\documentclass{ws-ijgmmp}

\usepackage{xcolor}

\newcommand{\be}{\begin{equation}}
\newcommand{\ee}{\end{equation}}
\newcommand{\bea}{\begin{eqnarray}}
\newcommand{\eea}{\end{eqnarray}}
\newcommand{\vsp}{\vspace{0.4cm}}

\newcommand{\calg}{\mathcal{A}}
\newcommand{\stsp}{\mathcal{D}_{\mathcal{A}}}
\newcommand{\obsp}{\mathcal{O}_{\mathcal{A}}}

\begin{document}

\markboth{Authors' Names}
{Instructions for Typing Manuscripts (Paper's Title)}

%
\catchline{}{}{}{}{}
%

\title{Differential calculus on manifolds with a boundary. Applications.}

\author{FLORIO MARIA CIAGLIA$^{1}$ $^{2}$, FABIO DI COSMO$^{1}$ $^{2}$, \\ MARCO LAUDATO$^{1}$ $^{2}$ $^{3}$ and GIUSEPPE MARMO$^{1}$ $^{2}$}

\address{$^{1}$ Dipartimento di Fisica ``E. Pancini'' dell’ Universit\'{a} ``Federico II'' di Napoli, Complesso Universitario di Monte S. Angelo, via Cintia, 80126 Naples, Italy.}

\address{$^{2}$ Sezione INFN di Napoli, Complesso Universitario di Monte S. Angelo, via Cintia, 80126 Naples, Italy.}

\address{$^{3}$ Turku Center for Quantum Physics, Department of Physics and Astronomy, University of Turku, FIN-20014, Turun yliopisto, Finland.}

\maketitle


\begin{abstract}
This paper contains a set of lecture notes on manifolds with boundary and corners, with particular attention to the space of quantum states.
A geometrically inspired way of dealing with these kind of manifolds is presented, and explicit examples are given in order to clearly illustrate the main ideas.

\end{abstract}


\tableofcontents

\section{Introduction} \label{section: introduction}

This paper contains a revised and extended version of the  lectures given by one of us (G.M.) at the ``International Workshop on Quantum Physics: Foundations and Applications'' held at the Centre for High
Energy Physics of the Indian Institute for Science of Bangalore. 
It contains a leisurely introduction to the differential calculus on manifolds with boundaries.
Its main motivation is differential calculus on the space of quantum states.
It deals with a geometrically inspired way of viewing and solving problems about subalgebras of smooth  functions and quotient manifolds.
The main idea is to investigate a manifold $M$ by means of the algebra of function $\mathcal{F}(M)$.
Actually, we shall argue that when it comes to the quotient manifolds and submanifolds it is convenient to consider also the full algebra of tensor fields on $M$, say $\mathcal{T}(M)$.
The differential aspects of the manifold are mainly carried by the Lie algebra of derivations (vector fields) of $\mathcal{F}(M)$.
The style  is rather informal. 
We emphasize examples, motivations and intuition and  refer to the literature for full proofs and the most
general definitions.

Manifolds with boundaries arise quite naturally in the description of realistic physical systems, both in classical and quantum physics. Typical examples are:

\begin{itemize}

\item Biliards, classical and quantum;

\item Electrons moving in wires, Hall effect;

\item Wave guides;

\item Source singularities.

\end{itemize}

The most simple example of a manifold with a boundary is the semiline $\mathbb{R}^{+}$, i.e., the real numbers $a\geq0$.
In general, the Cartesian product of manifolds with boundary is not a manifold with boundary, but a manifold with corners.
For instance, the Cartesian product of $N$ semilines is the prototype of manifold with corners.

In general, manifolds with boundaries arise every time we divide the carrier space into an accessible region and its complement which we assume to be not accessible to observation.
Thus, what happens on the boundary is an effective way to represent whatever happens ``on the other side''  which has an effect on the accessible part.

Most of the manifolds with a boundary we are going to be interested in arise as reduction of homogeneous spaces under  specific equivalence relations.
Thus, in principle we shall assume that manifolds with boundaries may be "unfolded" to manifolds without boundary.
As it is usually the case, the unfolding procedure is not uniquely defined, therefore, other considerations will enter the "unfolding" procedure, like symmetry considerations, minimality, simplicity, and so on.

In the context of $C^{*}$-algebras, this manifold may represent the space of states of a finite-dimensional, commutative $C^{*}$-algebra $\calg$.
Consequently, it represents the space of physical states of a finite-dimensional classical system.

In general, boundaries are generated by inequalities, the most common inequalities emerging in physical situations are associated with non negativity (energy, entropy production, probabilities and so forth).

To be a little more formal, we start by considering  the simplest carrier space, i.e., a vector space  $V$.
Let $\alpha\in V^{*}$ be an element in the dual space of $V$, and consider the map $\alpha:V\rightarrow\mathbb{R}$.
The inverse image of $\mathbb{R}_{+}$ is a half-plane, say $V^{+}$, and the inverse image of $0\in\mathbb{R}$ will be a hyperplane.
Using the same procedure with a set of linearly independent forms on $V$ we obtain what is known as a quadrant, or simplex.
The quadrant is the model space for the category of manifolds with corners (\cite{michor-manifolds_of_differentiable_mappings}).

Note that the procedure outlined above for finite-dimensional vector spaces makes sense also in the more general context of Banach Spaces.

Manifolds with corners very often arise as quotients by action of discrete group, that is, as orbifolds.
For instance, the half-plane may be considered also to be the orbifold of the vector space $V$ under the action of $\mathbb{Z}_{2}$, $(x\,;y)\mapsto(-x\,;y)$.
The quotient space would be diffeomorphic with $V^{+}$.
In intrinsic terms we would define an equivalence relation in $V$ by setting $v_{1}\sim v_{2}$ if $\left|\alpha(v_{1})\right|=\left|\alpha(v_{2})\right|$, i.e., either $\alpha(v_{1})=\alpha(v_{2})$ or $\alpha(v_{1})=-\alpha(v_{2})$.

This is interesting because, often, it is possible to ``double'' a manifold with a smooth boundary in order to unfold the quotienting procedure.
Consequently, we can discuss multi-differential operators and higher order differential operators directly on a manifold without a boundary and then apply a specific reduction procedure.
This should be kept in mind when discussing self-adjoint extensions and Dirac operator on manifolds with boundary.

\begin{figure}[h!]
\centering
\includegraphics[scale=0.3]{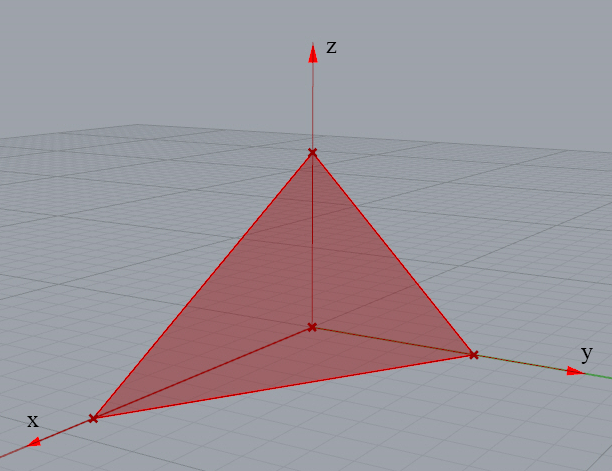}
\caption{\textit{Simplex representing the space of probability distributions of a three-level quantum system}}
\label{simplesso}
\end{figure}

Another interesting example comes from Quantum Mechanics.
The tomographic approach allows us to associate to every state $\rho$ a probability distribution $p_{\rho}$ with respect to a specific resolution of the identity.
For a finite level system, the space of probability distribution becomes an $n$-simplex, where $n$ is the dimension of the Hilbert space of the system (an example is shown in Fig.\ref{simplesso}).

This is clearly a manifold with corners.
The reader may consider the book \cite{bengtsson_zyczkowski-geometry_of_quantum_states:_an_introduction_to_quantum_entanglement} for additional aspects connected with Quantum Mechanics.

More generally, manifolds with boundary or corners are often associated with reduction procedures of dynamical systems, or actions of Lie groups or Lie algebras.
These procedures naturally lead to consider quotient spaces associated with dynamical systems. 
It could be the space of orbits or the space of leaves when more than one vector field is involved, for instance a family of vector fields closing on a Lie algebra.
Unfortunately, quotient spaces can be wild objects in the sense that they could lack of a differential structure as a smooth manifold, or, even worse, they could present pathological topologies, e.g., non-Hausdorff topologies.
In all these cases it would be difficult, if not impossible, to recover the notions of vector fields, differential forms, and tensor fields defined on the quotient space.

From a mathematical point of view, a convenient approach to deal with quotient spaces is the algebraic one, in which a manifold $\mathcal{M}$ is replaced by the algebra $\mathcal{F}(\mathcal{M})$ of functions on it.
Indeed, the algebra $\mathcal{F}(\mathcal{N})$ of functions on a quotient space $\mathcal{N}\cong\mathcal{M}/\sim$  can be identified with a subalgebra of the algebra of functions $\mathcal{F}(\mathcal{M})$ given by functions that are constant on the equivalence classes.
More generally,   when $\mathcal{N}$ admits a suitable differential structure, the algebra of covariant tensors on $\mathcal{N}$ can be represented as an invariant subalgebra (under the equivalence relation) of the space of covariant tensors on $\mathcal{M}$.
However this point of view can be used also in the cases in which $\mathcal{N}$ presents no differential structures. Indeed, in such a situation we are still able to define an invariant subalgebra of the algebra of tensor fields on $\mathcal{M}$ which we may declare to be the algebra of tensor fields on $\mathcal{N}$. Furthermore, this framework allows to replace the notion of vector field with the more general concept of derivation of an algebra, since vector fields on a smooth manifold are derivations of the algebra of smooth functions with respect to the pointwise product (\cite{willmore-the_definition_of_lie_derivative}).

In order to clarify these ideas, we will give some concrete examples.
Let us start with the semiline $\mathbb{R}^{+}_{0}$; it can arise by reduction from the real line $\mathbb{R}$ with respect to the action $x\mapsto -x$ of the discrete group $\mathbb{Z}_{2}$.
This space is the prototype of manifold with corners, and thus its differential structure is slightly more complex than that of a differential manifold.
From the algebraic point of view, this space can be described using the subalgebra $\mathcal{F}_{+}\subset\mathcal{F}(\mathbb{R})$ of ``even'' functions on the real line.
Not all the derivations of $\mathcal{F}(\mathbb{R})$ are derivations of $\mathcal{F}_{+}$. Indeed, since the derivative with respect to $x$ of an even function is an odd function, the module of derivations of $\mathcal{F}_{+}$ is generated by $x\,\frac{\partial }{\partial x}$.

For differential manifolds, we are used to the idea that, at least locally, all covariant tensors can be constructed out of functions and their differentials, but on the semiline, for instance, the tensor $\mathrm{d}x\otimes\mathrm{d}x$, globally well defined, cannot be generated by differentials of even functions.

A less trivial example comes from the two dimensional torus.
Specifically, let us consider $\mathbb{T}^{2}\cong S^{1} \times S^{1}$ and the vector field $X=a\,X_{\theta} + b \, X_{\phi}$, where $X_{\phi},X_{\theta}$ are the canonical vector fields on the circles composing the torus, and $a/b$ an irrational number.
It is well known that the quotient space with respect to this dynamical flow can be identified with the discrete set of irrational number, and the quotient topology on this set is then the discrete topology.
It is clear that there is no differential structure other than the trivial one.
However, writing $\alpha_{\theta},\alpha_{\phi}$ for the one forms that are dual to $X_{\theta},X_{\phi}$ respectively, we immediately find that $b\,\alpha_{\theta} - a\,\alpha_{\phi}$ is a basis for a one-dimensional module of differential one forms that are invariant with respect to the dynamical flow.
According to what has been said above, we could think of this module as the space of differential one forms on the quotient space.
The peculiar property is that the "quotient" would have forms and covariant tensors, but they would not be generated by differentials of functions defined on the quotient space.

A different pathology arises for instance with the space of orbits of the simple pendulum.
The phase space of this dynamical system is the cotangent bundle $T^{*}S^{1}$ over the manifold $S^{1}$. 
If one looks at the phase diagram, parametrized by energy and phase, it is possible to identify three different families of orbits: up to a certain value $E_{\star}$ of the energy $E$, the orbits are closed circles in the phase space; when $E=E_{\star}$ there are three orbits, one is a single point, and the other two are bounded but not closed; when $E>E_{\star}$ the orbits  can be divided into two classes according to the orientation of the actual motion.
The quotient space that arises from the reduction with respect to this dynamical flow is a non-Hausdorff space.

\begin{figure}[h!]
\centering
\includegraphics[scale=0.8]{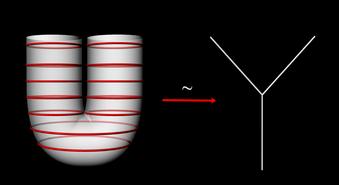}
\caption{\textit{Folding of the phase space $T^{*}S^{1}$ and its quotient with respect to the dynamics. For each value of the energy, represented in the z-direction, red curves are the corresponding dynamical orbits on $S^1$. It can be seen that, for $E > E_{\star}$, there are two different kinds of orbits: the orbits which are circulated clockwise, and the orbits which are circulated counterclockwise.}}
\label{pendolo}
\end{figure}



Apart from the absence of a differential structure, this space presents a topology with which it is difficult to work with.
However, from the algebraic point of view, the situation can be handled if we consider functions and forms which are invariant under the equivalence relation.
Since the energy function on the phase space is a constant of the motion, we can describe the quotient space by means of the invariant algebra $\mathcal{F}_{H}$, whose elements are all those functions in $\mathcal{F}(\mathcal{M})$ that depends on the energy.

Often, in the mathematical literature, when dealing with a manifold with a boundary, only vector fields which end up being tangent to the boundary are considered, what may be called the transversal part is usually assumed to vanish on the boundary.
The diffeomorphism group splits into the diffeomorphism group of the interior and the diffeomorphism group of the boundary.

\begin{figure}[h!]
\centering
\includegraphics[scale=0.3]{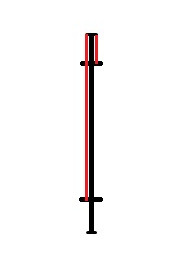}
\caption{\textit{Representations of the space of rotations around the $\hat{z}$-axis on the Bloch sphere. In red is drawn a gradient motion which moves the boundary towards the interior.}}
\label{segmento}
\end{figure}

From the physical point of view, however, we might be interested in the description of dynamical evolutions on manifolds with boundaries or corners.
This kind of dynamical evolutions often results in the motion from the boundary to the interior.
For example, let us consider the space of pure states of two-level quantum system (qubit).
This space can be seen as a two-dimensional sphere embedded into a three-dimensional Euclidean space.
Let us consider a rotation along the $\hat{z}$-axis.
The quotient space with respect to this dynamical flow can be identified with a closed interval of the real line (see Fig.\ref{segmento}), where the boundary points are the equivalence classes associated with the north and south poles respectively (fixed points).

\begin{figure}[h!]
\centering
\includegraphics[scale=0.6]{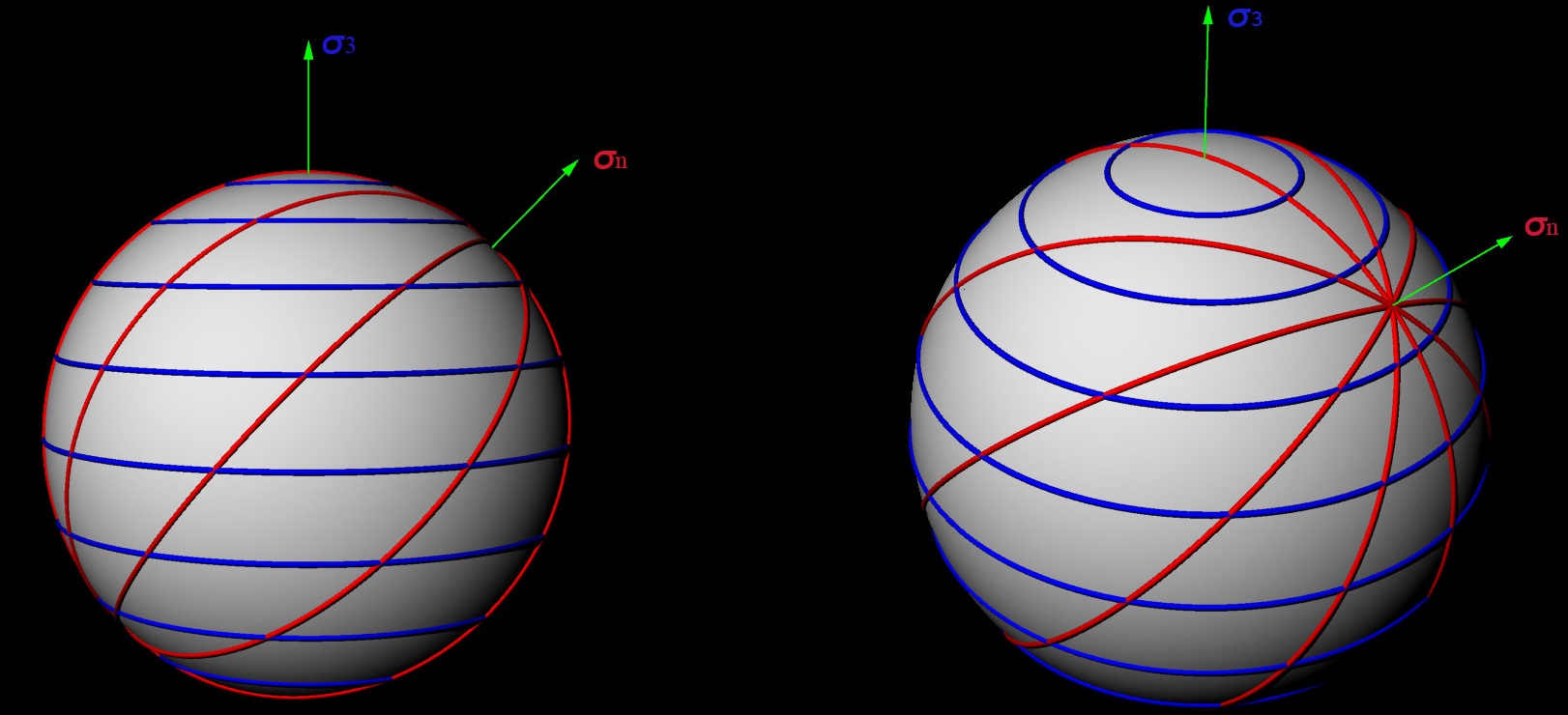}
\caption{\textit{Bloch sphere relative to a two-level quantum system. In green are represented rotations around the $\hat{z}$-axis; in red are drawn gradient motions with respect to $\hat{n}$-axis.  }}
\label{combo}
\end{figure}

Now, let us introduce a second axis $\hat{n}$ such that $\hat{n}\cdot\hat{z}\neq\pm 1$.
It is well known that, on the sphere, the so-called gradient motions\footnote{The Euclidean metric $\delta$ on the three-dimensional determines a rotationally-invariant metric $g$ on the sphere. 
The gradient motions are generated by vector fields $X_{f}$ such that $i_{X}g=df$ with $f$ a smooth function on the sphere.} with respect to the $\hat{n}$-axis are precisely the meridians on the sphere with respect to the $\hat{n}$-axis (see Fig.\ref{combo}). 
In particular, by considering a specific orbit  on the sphere, its projection would give rise to a motion with a fold; it is possible to move from the poles of the $\hat{z}$-axis, and thus, on the quotient space, the boundary points are mapped in the interior.

\vsp

Let us make some general considerations where we formalize some of the aspects we have been mentioning.

By means of the Whitney's theorem, manifolds may be always described as submanifolds of some $\mathbb R^N$, with $N$ sufficiently large.
In this respect, under mild regularity requirements, they may be thought of as arising from constraints:

\begin{equation}
\label{mar1}
f_1(m)=0,\;f_2(m)=0,\;\dots\;,f_k(m)=0.
\end{equation}

\noindent
Manifolds with a boundary arise, instead, from inequalites:

\begin{equation}
\label{mar2}
f_1(m)\geq0\;,f_2(m)\geq0,\;\dots\;,f_k(m)\geq0.
\end{equation}

\noindent
A simple example is a Ball:

\begin{equation}
\label{mar3}
x_1^2+x_2^2+\dots+x_k^2\leq1\qquad\text{or}\qquad1-(x_1^2+x_2^2+\dots+x_k^2)\geq0.
\end{equation}

\noindent
For $x_1,\,x_2,\,x_3$ this would be the Bloch ball for the quantum states of a q-bit. A billiard would be given by a submanifold in $\mathbb R^2$ defined by:

\begin{equation}
\label{mar4}
B=\{(x,y)\in\mathbb R^2\,|\,0\leq x\leq a,\,0\leq y\leq a\},\quad a\in\mathbb R.
\end{equation}

Manifolds with boundaries can be always thought of as an ``open part'' $\mathring{M}\subset M$ and a ``boundary part'' $\partial M\subset M$.
When there are corners, $\partial M$ is not a smooth submanifold.
When necessary, $M$ itself may be thought of as immersed in some open larger manifold of the same dimension. If we want to describe dynamical systems on these manifolds, we need to consider their tangent or cotangent bundles:

\begin{equation*}
\begin{split}
&TM\text{ for Newtonian and Lagrangian mechanics}\\
&T^*M\text{ for Hamiltonian mechanics.}
\end{split}
\end{equation*}

In his proof of the Atiyah, Patodi, Singer (APS) index theorem on manifolds with boundary, R. Melrose has developed a differential calculus and a b-geometry starting with an algebra of complete vector fields (\cite{melrose-introduction_to_microlocal_analysis}).
For instance, let us consider the $2$-dimensional example $\mathbb{R}\times\mathbb{R}^{+}_{0}$.
If for simplicity we introduce Cartesian orthogonal coordinates $(x\,;y)$ on $\mathbb{R}\times \mathbb{R}^{+}_{0}$, $y$ for $\mathbb{R}^{+}_{0}$ and $x$ for the transversal part to it, we can consider the algebra of vector fields $\left\{x\frac{\partial}{\partial x}\,,\frac{\partial}{\partial y}\right\}$.
These vector fields are complete and generate an action of an Abelian Lie group.
When multiplied by functions on $\mathbb{R}^{+}_{0}$ they define a Lie module.
On a manifold with corners we would have $x_{l}\frac{\partial}{\partial x_{l}}\;l\in\{1,...,k\}$, $\frac{\partial}{\partial y_{j}} \;j\in\{1,2,...,n-k\}$.

With the vector fields $x\frac{\partial}{\partial x}$ and $\frac{\partial}{\partial y}$, considered as derivations of $\mathcal{F}(\mathbb{R}^{+}_{0})$, we can construct the holomorph Lie algebra (\cite{carinena_ibort_marmo_morandi-geometry_from_dynamics_classical_and_quantum} page $523$).
Additionally, by noticing that with the similarity transformations $e^{-f}\left(x\frac{\partial}{\partial x}\right)e^{f}=x\frac{\partial}{\partial x} + x\frac{\partial f}{\partial x}$, and $\frac{\partial}{\partial y} + \frac{\partial f}{\partial y_{j}}$ we get first order differential operators out of derivations, we can construct the associative algebra of all differential operators.
Essentially, first order differential operators appear as a ``gauged'' version of vector fields (derivations), and we can either construct  multidifferential operators, or multivector fields.
There are two orbits of the Abelian group, one is the boundary and the other is the full interior.
The orbits generated by $x\frac{\partial}{\partial x}$ and $\frac{\partial}{\partial y}$ can never connect a point on the boundary with a point in the interior, that is, the diffeomorphisms generated by the algebra of vector fields preserve the boundary.

However, as we have already stressed, although this occurs often in the mathematical literature, this is unsatisfactory from the physical point of view.
In addition to all those vector fields that vanish on the boundary we need at least one that does not, otherwise we can not move from the boundary towards the interior\footnote{For instance, in the quantum case, this would mean that we are not able to describe dechoerence of a pure state, because dechoerence processess do not preserve the rank of quantum states.}.

It is clear that, in the case of a manifold with boundaries, the standard picture of the tangent bundle requires qualification.
Indeed, in this case we will have:

\begin{equation}
T\mathring{M}\text{ and }T(\partial M),\quad\text{or}\quad T^*\mathring{M}\text{ and }T^{*}(\partial M),
\end{equation}
and the sections of $T(\partial M)$, i.e., the vector fields tangent to the boundary, are not enough to implement the motion from the boundary to the interior.
To make clear this point let us introduce an example. 
Say $I\subset \mathbb R$ the interval:

\begin{equation}
I=\{x\in\mathbb R,\,|x|\leq1\}
\end{equation}

\noindent
which is considered as the configuration space for the dynamical system described by this Lagrangian function:

\begin{equation}
\label{mar5}
 L=\frac{1}{2}mv^2-\frac{1}{(1-x^2)^{\frac{1}{n}}},\quad n\in\mathbb N.
\end{equation}

\begin{figure}[h!]
\centering
\includegraphics[scale=0.38]{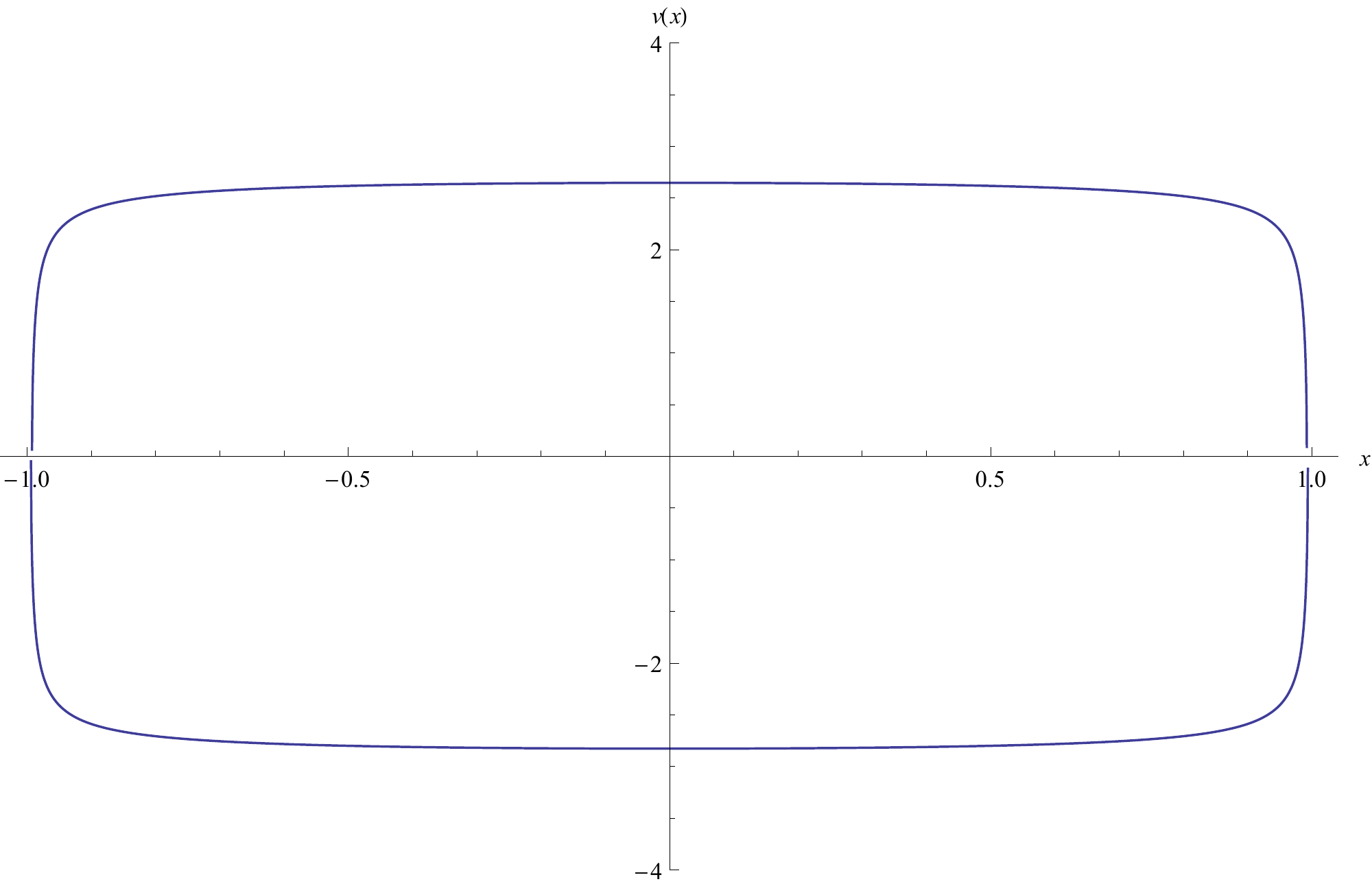}
\caption{\textit{For a given value of the total energy, in the limit $n\to\infty$, we would find that the velocity at the boundary would jump from $v$ to $-v$.}}
\label{fasespace}
\end{figure}

\begin{figure}[h!]
\centering
\includegraphics[scale=0.29]{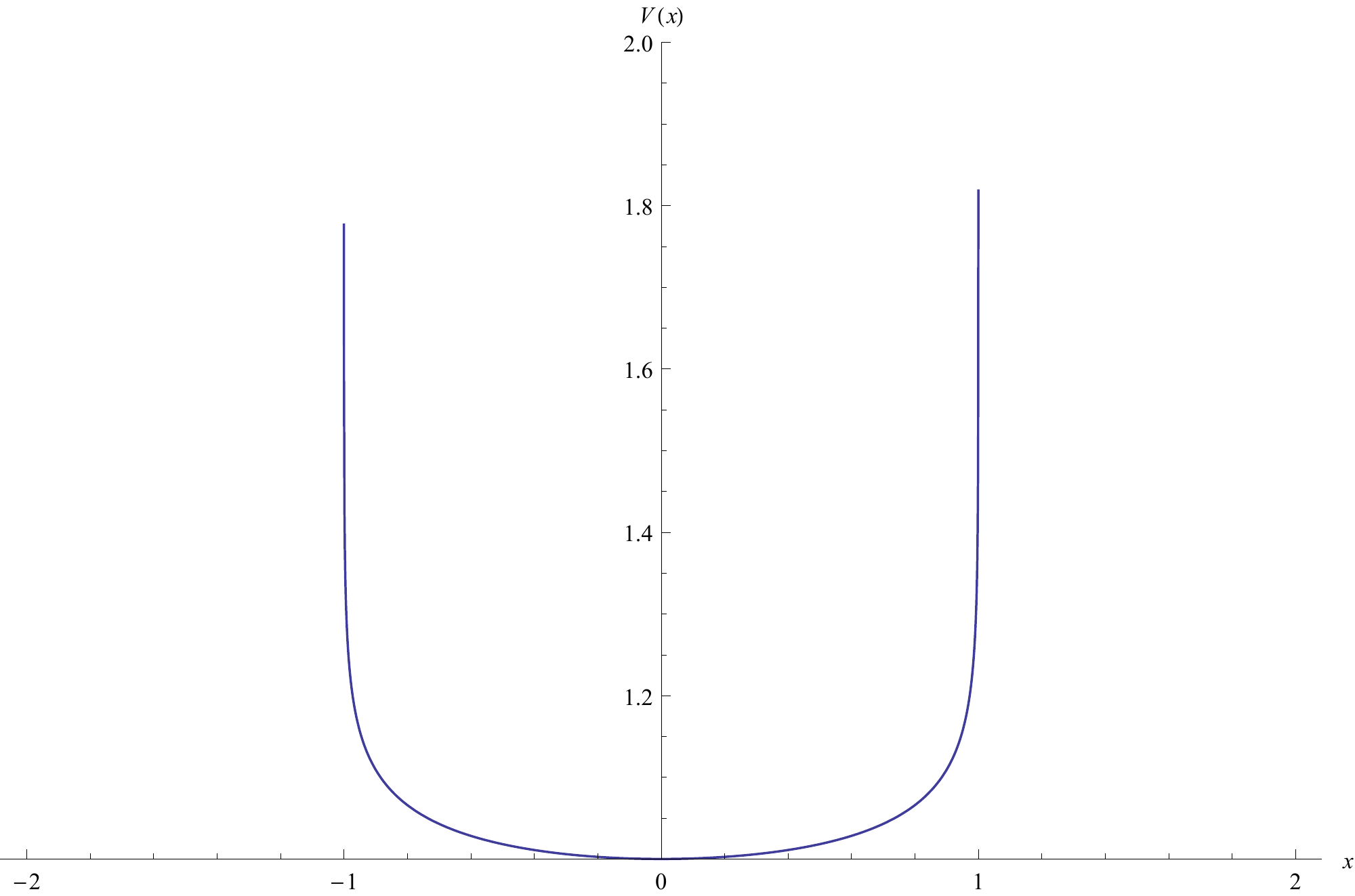}
\caption{\textit{In the limit $n\to\infty$ the dynamical system defined by Eq. (\ref{mar5}) will describe a free particle of mass $m$ which collides on the boundary and inverts its motion. Notice that the axes are centered in $(0,1)$}}
\label{n16}
\end{figure}

\noindent
If we now split the interval as $\mathring I$ and $\partial I$, the tangent bundles of these two manifolds have dimensions $dim\,T\mathring I=2$ and $dim\,T(\partial I)=0$. It means that the velocity at the boundary can be only zero and it is well accomodated by the phase portrait. However, if we want to consider the case in which $n\to\infty$, this structure is not enough. Indeed, in this limit, the potential becomes a "square well" (see Fig. \ref{n16}) and on the boundary all the values of the velocity are allowed (see Fig. \ref{fasespace}). To accomodate this situation, the fibers at the boundaries must have the same dimension they have at the interior.

A possible way to achieve this situation is by means of a pull-back bundle. Indeed, let us consider, in a more general fashion, an open manifold $\tilde Q$ which includes a manifold with boundary $Q$, namely $Q\subset\tilde Q$. 
Moreover, let us consider the vector bundle $E\overset{\pi}{\longrightarrow}\tilde Q$.
Then, by means of the immersion $\partial Q\overset{i}{\hookrightarrow} \tilde{Q}$, we can induce the pull-back bundle:

\begin{equation}
i^*(E(Q))=\{(q',e)\in\partial Q\times E\,|\,i(q')=\pi(e)\}\subset\partial Q\times E.
\end{equation}

\noindent
In other words, it is the union of the fibers:

\begin{equation}
\label{mar11}
\bigcup_{b\in\partial Q}E_b(Q).
\end{equation}

\noindent
which allows the boundary to have the same fiber of the inner points\footnote{Obviously, the same procedure can be carried on for the contangent bundle $T^*Q$ and momentum $p$.}.
Accordingly, the vectors of this fiber bundle are not all tangent to the boundary, there are some of them that allow to move the points from the boundary towards the interior.


\noindent
Returning to the example of the half-plane, we consider the vector field $\frac{\partial}{\partial x}$, which generates a semigroup of transformations on the half-plane in the sense that, in general, considering the generated flow $\{\phi_{\tau}\}$ on the ambient space, its general action would take us out of the half-plane, therefore, we have to restrict only to motions which take place in the half-plane.
With this restriction, the space of positions and allowed velocities will not be a manifold with boundary anymore.

\vsp

The paper is organized as follows.
First of all, we will review the basic tools of differential calculus starting with the case of a vector space $V$, and then passing to the more general case of a manifold without boundary.
By using the duality between a manifold and the algebra of smooth functions on it, we will outline the main features of differential calculus in an algebraic setting.
As shown above, this is extremely useful in dealing with unfolding and reduction procedures.
In particular, section \ref{section: vector spaces} will be devoted to natural constructions on vector spaces.
The algebraic formalism allows to immediately generalize the results of this section to the more general case of smooth manifolds by replacing scalars with scalar smooth functions.

The final step is enlarging the algebra of smooth functions to the exterior algebra of differential forms.
On this space it is possible to define a Clifford algebra structure by means of the so-called $\vee$-product.
This can be very useful in relation with the physics of spinors and Dirac operators.

Next, in section \ref{section: space of states} we will consider the space of states $\stsp$ of a finite-level quantum system associated with a finite-level $C^{*}$-algebra $\mathcal{A}$.
This is an example of a stratified manifold with boundary, that is, a topological space with boundary having no differential structure as a whole, but, rather, being made up of different strata each of which is a differential manifold on its own.
We will review the differential geometry of this space as it has been done in \cite{grabowski_kus_marmo-geometry_of_quantum_systems_density_states_and_entanglement}.
The space $\stsp$ will always be thought of as a subset of a linear space, namely, the dual space $\obsp^{*}$ of the space of observables, so that a global differential structure in some ambient space will always be at our disposal.
The differential structure of different strata will be, essentially, the differential structure inherited from this ambient space.
According to the ideology exposed in this introduction, this allows us to bypass some of the technical problems due to the absence of a global differential structure on $\stsp$. 
Consequently, we will show how to implement the notion of a vector field $\Gamma_{L}$ which is transversal to the strata of $\stsp$, so that we are be able to describe some interesting physical phenomena that do not preserve the rank of quantum states, e.g., dechoerence or damping phenomena.
This vector field $\Gamma_{L}$ is defined on $\obsp^{*}$ and the fact that it is transversal to the strata of $\stsp$ is implemented  by looking at its action, as a derivation of $\mathcal{F}(\obsp^{*})$, on the so-called expectation value functions.

\section{Natural Constructions and Differential Calculus on Vector Spaces}\label{section: vector spaces}

Before embarking in the differential calculus on manifolds with a boundary it is convenient to review differential calculus on manifolds without boundary. 
The advent of non-commutative geometry in physics has stressed the role of the identification of a manifold with the algebra of functions defined on it:

\begin{equation}
\label{mar12}
\mathbb M\rightleftharpoons\mathcal F(\mathbb M).
\end{equation}

\noindent
Where $\mathcal F$ is given, $\mathbb M$ can be recovered as the space of algebra homomorphisms $\text{Hom}_{\mathbb K}(\mathcal F,\mathbb K)$ with $\mathbb K$ the field of real or complex numbers. 
For further details we refer to \cite{marmo_morandi-some_geometry_and_topology, landi_marmo-algebraic_differential_calculus_for_gauge_theories} and \cite{nelson-tensor_analysis}.

It is useful to step back and recall that $\mathbb M$ may be thought of as a generalization of a vector space $V$, it means that for each $m\in\mathbb M$, there is an open neighbourhood of $m$, say $U_m$, diffeomorphic to $V$ or, said differently, in one-to-one correspondence in a smooth way to $V$ (or an open contractible neighbourhood of the origin). 
Thus, for simplicity, we may consider $\mathbb M$ to be a vector space $V$ or an affine space having $V$ as a model. 
All geometrical or algebraic structures we may build out of $V$, without any further requirements, may also be exported on a generic manifold when they are glued together as

\begin{equation}
\label{mar13}
\bigcup_{m\in\mathbb M}V_m.
\end{equation}

We first notice that along with $V$ we have $V^*\equiv\text{Lin}_{\mathbb K}(V,\mathbb K)$, this is the space of linear functions on $V$, being functions they can be multiplied point-wise and provide us with an associative, commutative algebra with identity, say $\mathcal F(V)$, which contains all polynomials built out of $V^*$.
These linear functions may also be multiplied differently, the pointwise product is:

\be
(f\cdot g)(v):=f(v)g(v)\,,
\ee 
but also as providing a function to be evaluated on pair of vectors:

\be
(f\times g)(u\,,v):=f(u)g(v)\,.
\ee
Similarly we can build multilinear maps.
If we have in mind the transition to manifolds, we should allow for general nonlinear transformations, not restricting only to linear transformations.
A tensorial description of the vector space structure of $V$ is provided by the homogeneous first order differential operator $\mathcal E$, called Euler operator, with the following properties:

\begin{itemize}

\item[i)] There exists only one point $v_0\in V$ such that $\mathcal E(v_0)=0$;

\item[ii)] $\mathcal E$ is a complete vector field, its flow defines a one-parameter group of transformations;

\item[iii)] $\mathcal E\,f=1\cdot f$ has $n$ independent solutions if $n=\text{dim }V$;

\item[iv)] $\mathcal E\,f=0\cdot f$ has only constant solutions.

\end{itemize}

\noindent

Notice that conditions ii), iii) and iv) would be satisfied on some open neighborhood of $v_{0}$, therefore the required completeness  implies that the open neighborhood should actually give rise to the full vector space.

By means of the third requirement, we can select $n$ independent linear functions and use them to define the vector space structure. 
Notice that here linear independence coincides with functional independence.

It is clear that $\mathcal F(V)$, considered as an algebra of functions, is defined by:

\begin{equation}
\label{mar16}
\begin{split}
&(f_1+f_2)(v)=f_1(v)+f_2(v)\\
&(f_1\cdot f_2)(v)=f_1(v)f_2(v)\\
&h\,(f_1+f_2)=h\cdot f_1+h\cdot f_2
\end{split}
\end{equation}

\noindent
and therefore contains no reference to the vector space structure of $V$. 
This information will be carried by the Euler operator $\mathcal E$ which allows to define homogeneous polynomial functions of degree $k$ by means of the following eigenvalue equation:

\begin{equation}
\label{mar15}
\mathcal E(f)=k\,f.
\end{equation}

\noindent
Since this  characterization is given in terms of a vector field, this description does not depend on the choice of a particular coordinate system. 
Indeed, in any linear coordinate system\footnote{Say $(e_1,\,e_2,\dots,e_n)$ a basis of solutions of iii), we define the dual basis $(e^1,\,e^2,\dots,e^n)$ such that $e^j(e_k)=\delta^j_k$, and we set $x^j(v)=e^j(v)$, $v\in V$.} for $V$ the first order differential operator is:

\begin{equation}
\label{mar14}
\mathcal{E}=x^j\frac{\partial}{\partial x^j}.
\end{equation}

\noindent
If we now select a different Euler operator, for instance $\mathcal E=(x^j+c^j)\frac{\partial}{\partial x^j}$, it will have all the properties of a "linear structure" and therefore a new polynomial subalgebra of $\mathcal F(V)$. From this point of view $\mathcal F(V)$ has some "universal properties" because it only depends on the differential properties of $V$ and does not depend on its linear structure (see \cite{carinena_ibort_marmo_morandi-geometry_from_dynamics_classical_and_quantum}).

\subsection{Tensors on a vector space and on its dual}
In finite dimensions, $V$ coincides also with the linear function defined on $V^*$:

\begin{equation}
\label{mar17}
V=\text{Lin}_\mathbb{K}(V^*,\mathbb{K}),\quad\text{i.e.,}\quad V=(V^*)^*.
\end{equation}

\noindent
In infinite dimensions this relation is not guaranteed and when it holds we say that $V$ is totally reflexive.

From linear maps\footnote{From now on we will assume that all the vector spaces are defined on the same common field $\mathbb{K}$, and thus, for the sake of notational simplicity, we will drop the indication of $\mathbb{K}$.} $\text{Lin}(V)$ we can generalize to multilinear maps:

\begin{equation}
\label{mar18}
\text{MLin }(V\times V\times\dots\times V)
\end{equation}

\noindent
and dually:

\begin{equation}
\label{mar19}
\text{MLin }(V^*\times V^*\times\dots\times V^*)
\end{equation}

\noindent
along with multilinear mixed maps:

\begin{equation}
\label{mar20}
\text{MLin }(V\times\dots\times V\times V^*\times\dots\times V^*).
\end{equation}

\begin{remark}
We should pay attention to the fact that multinear maps are not linear maps.
\end{remark}

\noindent
These multilinear maps are also called tensors and we denote them by:

\begin{equation}
\label{mar21}
\mathcal T^\sigma(V),\quad\mathcal T^\sigma(V)\cong\mathcal T_\sigma(V^*)
\end{equation}

\noindent
or, in the mixed case, $\mathcal T^\sigma_\rho(V)$. We denote by:

\begin{equation}
\label{mar22}
\mathcal T^0(V)\equiv\mathbb K,\quad\mathcal T'(V)=V^*.
\end{equation}

\noindent
We shall also write as a product, say:

\begin{equation}
\label{mar23}
\begin{split}
&\text{MLin }(V\times V)\equiv V^*\otimes V^*\\
&\text{MLin }(V\times V\times V)\equiv V^*\otimes V^*\otimes V^*=(V^*\otimes V^*)\otimes V^*
\end{split}
\end{equation}

\noindent
and so on.

The tensor product $\otimes$ defines an associative algebra structure on tensors. It contains two subalgebras of totally symmetric tensors and totally skew-symmetric tensors. Occasionally, we shall denote them by $\mathcal \mathcal{S}(V^*)$ and $\Lambda(V^*)$.

\subsection{Vector spaces with additional structures}

Assume $V$ carries a skew-symmetric binary, bilinear product denoted by a square bracket:

\begin{equation}
\label{mar24}
[\,,\,]\,:\,V\times V\longrightarrow V
\end{equation}

\noindent
such that:

\begin{itemize}

\item[i)] $[v_1,v_2]=-[v_2,v_1]$

\item[ii)] $\bigl[v_1,[v_2,v_3]\bigr]=\bigl[[v_1,v_2],v_3\bigr]+\bigl[v_2,[v_1,v_3]\bigr]$.

\end{itemize}

\noindent
This pairing defines a Lie algebra structure. Property ii) says that this operation defines derivations for the binary, bilinear product. 
It is usually called the Jacobi identity.

It is possible to develop an exterior differential calculus at a purely algebraic level. When in addition the Lie algebra $(V,[\,,\,])$ acts as an algebra of derivations on $\mathcal F(\mathbb N)$, with $\mathbb N$ any smooth manifold, the algebraic differential calculus becomes richer and contains all essential ingredients of the usual differential calculus on manifolds.

In the few coming sections we shall illustrate these various aspects. 
Before entering these aspects, let us remark that any Lie algebra structure on $V$ induces a Poisson bracket on $\mathcal F(V^*)$. 
This is done first on linear functions by exploiting the identification $V\equiv\text{Lin }(V^*)$ and then extended by using the derivation property with respect to the associative, commutative product on the point-wise algebra $\mathcal F(V)$. We have, if we denote by $\hat v$ the linear function corresponding to $v$

\begin{equation}
\label{mar25}
\{\hat v,\hat u\}=\widehat{[v,u]}
\end{equation}

\noindent
then 

\begin{equation}
\label{mar26}
\{\hat v,\hat u\hat w\}=\{\hat v,\hat u\}\hat w+\hat u\{\hat v,\hat w\}.
\end{equation}

\noindent
As $\hat v$ and $\hat u$ are (linear) functions, we can define also a controvariant tensor field on $V^*$ by setting:

\begin{equation}
\label{mar27}
\Lambda(d\hat v,d\hat u)(\alpha)=\alpha\bigl([v,u]\bigr)
\end{equation}

\noindent
where $\alpha\in V^*$.

This bivector field satisfies a homology property when applied to higher multivectors, it is derived from the Jacobi identity, and is the counterpart of $d\cdot d=0$ on forms (\cite{landi_marmo-algebraic_differential_calculus_for_gauge_theories}).

\subsection{Homology Operators on Multivectors}

We have (\cite{koszul-homologie_et_cohomologie_des_algebres_de_lie}) a sequence:

\begin{equation}
\label{mar28}
\Lambda^p(V)\overset{\partial}{\longrightarrow}\Lambda^{p-1}(V)\overset{\partial}{\longrightarrow}\dots\overset{\partial}{\longrightarrow}\Lambda^1(V)\longrightarrow0
\end{equation}

\noindent
defined by:

\begin{equation}
\label{mar29}
\partial(v_1\wedge v_2\wedge\dots\wedge v_p)=\sum_{i<j}(-1)^{i+j+1}[v_i,v_j]\wedge v_1\wedge\dots\wedge \hat{v_i}\wedge\dots\wedge \hat{v_j}\wedge\dots\wedge v_p\,,
\end{equation}
where the $\hat{}$ notation means that the corresponding vector is missing from the expression.

\noindent
It follows that $\partial\cdot\partial=0$. We remark that $\partial$ is not a derivation with respect to the wedge product. We have also

\begin{equation}
\label{mar30}
\epsilon_v\,:\,\Lambda^p\longrightarrow\Lambda^{p+1}
\end{equation}

\noindent
by:

\begin{equation}
\label{mar31}
\epsilon_v\colon v_1\wedge v_2\wedge\dots\wedge v_p\longmapsto v\wedge v_1\wedge v_2\wedge\dots\wedge v_p
\end{equation}

\noindent
with the property 

\begin{equation}
\label{mar32}
\epsilon_v\epsilon_u=-\epsilon_u\epsilon_v.
\end{equation}

\noindent
It is now possible to define:

\begin{equation}
\label{mar33}
L_u=\epsilon_u\partial+\partial\epsilon_u
\end{equation}

\noindent
which turns out to be the extension to $\Lambda(V)$ of the adjoint representation

\begin{equation}
\label{mar34}
ad\,:\,V\longrightarrow \text{Lin }(V,V)
\end{equation}

\noindent
by:

\begin{equation}
\label{mar35}
v\longmapsto ad_v
\end{equation}

\noindent
with $ad_v(u)=[v,u]$.

Given two multivectors $G\in\Lambda^q(V)$, $H\in\Lambda^p(V)$ we define the extension of the Lie algebra product on $V$ to the algebra of multivectors by setting:

\begin{equation}
\label{mar36}
(-1)^{q+1}[G,H]=\partial(G\wedge H)-\partial G\wedge H-(-1)^qG\wedge\partial H.
\end{equation}

\noindent
This bracket is the Schouten-Nijenhuis bracket (\cite{frolicher_nijenhuis-theory_of_vector_valued_differential_forms_I, frolicher_nijenhuis-theory_of_vector_valued_differential_forms_II}) which makes $\Lambda(V)$ a graded Lie algebra. We should remark that the r.h.s. appears as the ``deviation'' from a graded derivation of $\partial$.

Given any endomorphism (\cite{marmo_ferrario_lovecchio_morandi_rubano-the_inverse_problem_in_the_calculus_of_variations_and_the_geometry_of_the_tangent_bundle})

\begin{equation}
\label{mar37}
T\,:\,V\longrightarrow V
\end{equation}

\noindent
we consider the natural extension to $\Lambda(V)$ by setting

\begin{equation}
\label{mar38}
\delta_T\,:\,v_1\wedge v_2\wedge\dots\wedge v_p\longmapsto\sum_j v_1\wedge v_2\wedge\dots\wedge T(v_j)\wedge\dots\wedge v_p.
\end{equation}

\noindent
For $T=\mathbf{1}$, we simply get:

\begin{equation}
\label{mar38}
\delta_{1}\,:\,v_1\wedge v_2\wedge\dots\wedge v_p\longmapsto p(v_1\wedge v_2\wedge\dots\wedge v_p).
\end{equation}

\noindent
Now we combine these operations to get:

\begin{equation}
\label{mar39}
\partial_T=\delta_T\cdot\partial-\partial\cdot\delta_T\,:\,\Lambda(V)\longrightarrow\Lambda(V).
\end{equation}

\noindent
In general $\partial_T\cdot\partial_T\neq0$ and we have:

\begin{equation}
\label{mar40}
\partial_T(\partial_T(v_1\wedge v_2\wedge v_3))=N_T(v_1,v_2,v_3)\in V
\end{equation}

\noindent
defines a tensor associated with $T$ which we might call a Nijenhuis tensor. 
This tensor vanishes iff $\partial_T\cdot\partial_T=0$ and is a boundary homology operator. Now we want to dualize this construction. We first set for any linear function on $V$, say $\alpha\in V^*$,

\begin{equation}
\label{mar41}
d\alpha(v_1\wedge v_2)=\alpha\bigl([v_1,v_2]\bigr).
\end{equation}

\noindent
This time $d\cdot d=0$ because of the Jacobi identity and on a generic antisymmetric multilinear map we get

\begin{equation}
\label{mar42}
d\beta(v_1\wedge v_2\wedge\dots\wedge v_{p+1})=\sum_{i<j}(-1)^{i+j}\beta\bigl([v_i,v_j]\wedge v_1\wedge\dots\wedge v_i\wedge\dots\wedge v_j\wedge\dots\wedge_{p+1}\bigr).
\end{equation}

\noindent
Again we define a contraction

\begin{equation}
\label{mar43}
i_v\,:\,\Omega^p\longrightarrow\Omega^{p-1},\qquad i_v\Omega^0=0
\end{equation}

\noindent
with the following properties:

\begin{equation}
\label{mar44}
i_vi_u+i_ui_v=0;\qquad i_vd+di_v=L_v\,.
\end{equation}

\noindent
$L_{v}$ will be called the Lie derivative and we get:

\begin{equation}
\label{mar45}
[i_vd+di_v,i_u]=i_{[v,u]}.
\end{equation}

\noindent
Moreover

\begin{equation}
\label{mar46}
[L_u,L_v]=L_{[u,v].}
\end{equation}

\noindent
We may now define pairings on $V\otimes V^*$ by setting:

\begin{equation}
\label{mar47}
\bigl<(v,\alpha)|(u,\beta)\bigr>_+=\alpha(u)+\beta(v)
\end{equation}

\noindent
or

\begin{equation}
\label{mar48}
\bigl<(v,\alpha)|(u,\beta)\bigr>_-=\alpha(u)-\beta(v).
\end{equation}

\noindent
It is possible to define a pairing between $\Lambda(V^*)$ and $\Lambda(V)$ by setting:

\begin{equation}
\label{mar49}
\bigl<v_1\wedge v_2\wedge\dots\wedge v_p|\alpha^1\wedge\alpha^2\wedge\dots\wedge \alpha^p\bigr>=det\,||\alpha^j({v_k})||.
\end{equation}

\noindent
With respect to this pairing $d$ and $-\partial$ are the transpose one of the other, $i_v$ is the transpose of $\epsilon_v$.

When $V$ itself carries an inner product, it is possible to induce an inner product on the associated tensor spaces.
It is also possible to construct an Hodge-$*$ operator which allows to build second order ``differential operators'' out of the introduced exterior derivative.
In particular this would give a Laplacian or a D'Alambertian operator when we move from vector spaces to modules and the Lie algebra is realized in terms of vector fields.
In the next section we will make this statement more precise.

In this section we have shown how to construct differential calculus at purely algebraic level, in the sense that there was no need to introduce the concepts of points, limits, and quotients to define derivatives.

\section{Manifolds}

It is possible to move all our previous constructions from a vector space V to a manifold $\mathbb{M}$. Essentially, each contractible open neighbourhood of a point will be diffeomorphic to its tangent space, therefore, all of our constructions are glued together by taking the union over points of M. 
With this construction all tensors become tensor fields, the field of coefficients, real or complex numbers will be replaced by the ring $\mathcal{F}(\mathbb{M})$ of functions on $\mathbb{M}$, either real or complex valued.\newline
The group $\mathrm{Diff}(\mathbb{M})$ for any differential manifold says that all points look alike. 
Now we can construct $T\mathbb{M}$, $T^*\mathbb{M}$, $\mathcal{T} ^r_s(\mathbb{M})$, and the jet bundle $J^k(\mathbb{M})$. 
A dualization, replacing $\mathbb{M}$ with $\mathcal{F}(\mathbb{M})$, is possible, sections of $T\mathbb{M}$ will be replaced by $\mathrm{Der}(\mathcal{F}(\mathbb{M}))$, the derivations of $\mathcal{F}(\mathbb{M})$. 
It is possible to consider $\mathcal{F}(\mathbb{M})$ as a trivial Abelian Lie algebra and construct the Holomorph of $\mathcal{F}$ by considering the semidirect product $\mathrm{Der}\mathcal{F}\ltimes \mathcal{F}$, $\left[ (D_1,f_1), (D_2,f_2) \right]=(\left[D_1,D_2\right],D_1f_2-D_2f_1)$. 
If $\mathcal{F}$ is thought of as an Abelian group we can consider $\mathrm{Aut}(\mathcal{F})\times \mathcal{F}$. This semidirect product is a very large infinite dimensional group which contains the semidirect extension $\mathrm{Diff}(\mathcal{F})\ltimes \mathcal{F}$ as a subgroup, that is the subgroup of automorphism which not only respect the vector space structure of $\mathcal{F}$, but also the pointwise associative and commutative product on $\mathcal{F}$. \newline
As an example of an automorphism of $\mathcal{F}$ as an Abelian group which is not an automorphism of $\mathcal{F}$ as an algebra we can consider 
$$
e^{-it\Delta} : L^2(\mathbb{M})\rightarrow L^2(\mathbb{M})
$$  
i.e., we consider the subspace of square integrable functions on $\mathbb{M}$ as an Abelian vector group. Any self-adjoint operator will give rise to a one-parameter group of transformations which will never respect the algebra structure, unless it is described by a first order homogeneous differential operator. \newline
The enveloping algebra of $Hol(\mathcal{F})$, the holomorph of the Abelian Lie algebra $\mathcal{F}$, represents the algebra of differential operators acting on $\mathcal{F}$. \newline
These remarks allow to state that by considering several copies of $\mathcal{F}$ we could define Multi-differential operators by means of the action on functions
$$
(D_1\otimes D_2\otimes \cdots\otimes D_n)(f_1,f_2,\cdots,f_n)=D_1f_1 \cdot D_2f_2 \cdots D_nf_n
$$   
It is also possible to replace first order differential operators with derivations, say multi-derivation maps 
$$
X_1\otimes X_2 \otimes \cdots \otimes X_n
$$
Out of these multi-operators it is possible to define higher order differential operators when an ordering is defined. For instance one could write $D_1(D_2(\cdots D_n f))$ or $X_1(X_2\cdots X_nf)$. The differential operator we obtain in this manner suffers of the same problems we have when defining a ``quantization procedure'' for polynomial functions in the momenta, i.e., we have ordering problems.

\begin{remark}
It is worth spending at this point some more words on the duality between vector fields and derivations of the Abelian algebra $\mathcal{F}(\mathbb{M})$. 
Even if Gelfand-Naimark theorem states that every commutative $C^*$-Algebra can be represented isometrically as the algebra $C^0(\mathbb{M})$ of continuous functions on a suitable manifold $\mathbb{M}$, it does not says that every commutative $C^*$-Algebra is the algebra $C^0(\mathbb{M})$. 
This means that the identification between vector fields and derivations of the algebra of functions cannot be based on the commutativity alone. 
Derivations of the algebra $\mathcal{F}(\mathbb{M})$ of functions on the manifold $\mathbb{M}$, indeed, are vector fields if and only if the product of two functions is local, i.e. it does not increase the support of the two functions. In order to make clearer this point let us consider a simple example. \newline
Let $\mathbb{R}$ the real line and $\mathcal{F}(\mathbb{R})$ the algebra of functions on $\mathbb{R}$ equipped with the Abelian product:
$$
(f\star g) (p) = \int_{-\infty}^{+\infty} f(p-q)g(q)dq
$$
that is the usual convolution product. As it is easily seen the support $\mathrm{supp}(f\star g)$ of the element $f\star g$ can be grater of the intersection $\mathrm{supp}(f) \cap \mathrm{supp}(g)$: this means that the product is not local. \newline
The multiplication by $p$ is a derivation of this product as shown in the following:
$$
(p(f\star g))(p)=p \int_{-\infty}^{+\infty} f(p-q)g(q)dq = 
$$
\begin{equation*}
\int_{-\infty}^{+\infty} ((p-q)f(p-q)g(q)+f(p-q)qg(q))dq =((pf)\star g)(p)+ (f\star (pg))(p)
\end{equation*}
Clearly this is not a vector field. 
However by means of Fourier transform $\phi \, : \, \mathcal{F}(\mathbb{R}) \, \rightarrow \, \mathcal{F}(\mathbb{R})$, we map $\mathcal{F}(\mathbb{R})$ with the convolution product, into the algebra $\mathcal{F}(\mathbb{R})$ equipped with the usual pointwise product $\cdot$. It is well known, indeed, that 
$$
\phi((f\star g)(p))= \phi(f)(x)\cdot \phi(g)(x)
$$  
and the derivations of this product are the usual vector fields $X=A(x)\frac{\partial}{\partial x}$. Therefore this shows that the same commutative $C^*$-Algebra (the $C^*$-Algebra of multiplicative operator on the Hilbert space $L^2(\mathbb{R})$) is represented as the pointwise algebra of continuous functions on $\mathbb{R}$ only in the second case. 
\end{remark}

\begin{remark}
We should point out that there could be local products which are associative and commutative, whose derivations are vector fields, but they do not generate the tangent space at each point of the manifold.
This is the case for $f_{1} \cdot_{K} f_{2}=f_{1}\cdot K\cdot f_{2}$ with $K$ any function on the manifold.
Clearly, derivations of the deformed product will be all the vector fields which admit $K$ as a constant of the motion, in particular, if $dim(M)=n$,at point of the manifold they will generate a vector space of dimensions $n-1$.
\end{remark}

\subsection{Brackets}

Following \cite{grabowski-brackets}, the algebra of differential operators may be considered an associative algebra with bilinear operation given by the composition 
$$
(D_1\cdot D_2)(f)= D_1(D_2f)\,.
$$
It contains the subalgebra of zeroth order operators (multiplication operators). It may be compared with the space of all linear endomorphism $End(\mathcal{F})$.
Out of this binary operation it is possible to produce another bilinear binary operation, namely the commutator 
$$
\left[ D_1,D_2 \right] = D_1\cdot D_2 - D_2 \cdot D_1
$$
and the anticommutator 
$$
(D_1,D_2)= D_1 \cdot D_2 + D_2 \cdot D_1
$$
The first bracket defines a Lie algebra structure, the second one a Jordan algebra structure. 
Of course we have the Jacobi identity 
$$
\left[ D, \left[ D_1, D_2 \right]\right]= \left[\left[ D, D_1 \right],D_2\right]+\left[ D_1, \left[ D, D_2 \right]\right]
$$
and 
$$
\left[ D, \left( D_1, D_2 \right)\right]= \left(\left[ D, D_1 \right],D_2\right)+\left( D_1, \left[ D, D_2 \right]\right)
$$
It is quite remarkable that we have the identity 
$$
(D_1,(D_2,D_3))-((D_1,D_2),D_3)=\left[D_1,\left[ D_2,D_3\right]\right]-\left[\left[D_1,D_2\right],D_3\right]
$$
i.e., they define a Lie-Jordan algebra. \newline
We recall the definition:\newline
A Lie bracket on an associative algebra such that the Leibnitz rule 
$$
\left[ D, D_1\cdot D_2\right]= \left[ D, D_1 \right] \cdot D_2 + D_1 \cdot \left[ D, D_2 \right] 
$$ 
is satisfied is called a Poisson bracket and the triple is called a Poisson algebra.
According to Dirac, it would be a q-Poisson bracket defining a q-Poisson algebra (\cite{Emch-mathematical_and_conceptual_foundations_of_20th_century_physics, landsman-between_classical_and_quantum}). \newline
Clearly when the associative algebra is commutative, the commutator bracket vanishes therefore the Poisson structure is an extra structure made out of outer derivations (instead of inner derivations). 
In general a Poisson structure is associated with a bivector field $\Lambda$, say
$$
\left\lbrace f,g  \right\rbrace = \Lambda(df,dg)
$$
with $$\Lambda = \frac{1}{2}\Lambda^{jk}\frac{\partial}{\partial x^j}\wedge \frac{\partial}{\partial x^k}$$
and 
$$
\Lambda(df,dg)=\Lambda^{jk}\frac{\partial f}{\partial x^j}\wedge \frac{\partial g}{\partial x^k}
$$
The Jacobi identity requires 
$$
\sum_i \left( \Lambda^{ij}\frac{\partial \Lambda^{kl}}{\partial x^i} + \Lambda^{ik}\frac{\partial \Lambda^{lj}}{\partial x^i} + \Lambda^{il}\frac{\partial \Lambda^{jk}}{\partial x^i} \right)=0
$$
\begin{remark}
For any differential operator D of order k on $\mathbb{M}$ we define the principal symbol to be the multilinear map
$$
\sigma(D)(df_1,df_2, \cdots, df_k)= \left[\cdots \left[D,\hat{f}_1\right],\hat{f}_2, \cdots \hat{f}_k\right]
$$
where $\hat{f}$ is the zeroth order multiplication operator. By using the Lie algebra structure of the algebra of differential operators we find that $\sigma(D)$ is a symmetric k-multilinear map which defines a k-polynomial function along the fibers of $T^*\mathbb{M}$. 
In this manner, by using the canonical Poisson bracket on $T^*\mathbb{M}$ it is possible to define a bracket on symbols, say
$$
\left\lbrace \sigma(D_1),\sigma(D_2) \right\rbrace = \sigma(\left[ D_1,D_2 \right])
$$
\end{remark}

\begin{example}
A Jacobi bracket on the space $\mathcal{F}(\mathbb{M})$ of functions on a manifold $\mathbb{M}$ is a bracket $\left\lbrace \cdot\,,\cdot \right\rbrace_{(\Lambda,\Gamma)}$ associated with a pair $(\Lambda, \Gamma)$, where $\Lambda$ is a bivector field, and $\Gamma$ is a vector field.
It must satisfy the Jacobi identity, which, in this case, can be translated in the requirement  that 
$$
\left[ \Lambda, \Lambda \right]=2\Lambda \wedge \Gamma \; , \qquad L_{\Gamma}\Lambda=0
$$
so that the bracket may be written in the form
$$
\left\lbrace f,g \right\rbrace_{(\Lambda,\Gamma)}=\Lambda(df,dg)+(L_{\Gamma}f)g-fL_{\Gamma}g\,.
$$
This bracket does not satisfy the Leibnitz rule and $\left\lbrace f, \cdot \right\rbrace_{(\Lambda,\Gamma)}$ acts on $C^{\infty}(M)$ as a first order differential operator not as a derivation. 
Indeed we have
$$
\left\lbrace f,gh \right\rbrace_{(\Lambda,\Gamma)} = \left\lbrace f, g \right\rbrace_{(\Lambda,\Gamma)} h + g \left\lbrace f,h \right\rbrace_{(\Lambda,\Gamma)} - \left\lbrace f,1 \right\rbrace_{(\Lambda,\Gamma)}gh \, . 
$$

For instance, on a contact manifold $(\mathbb{M},\theta)$ we can define $\Lambda$ and $\Gamma$ by means of the expressions:
$$
i_{\Lambda}\left(\theta\wedge(\mathrm{d}\theta)^{n}\right)=n\,\theta\wedge(\mathrm{d}\theta)^{n-1}\,,
$$
$$
i_{\Gamma}\left(\theta\wedge(\mathrm{d}\theta)^{n}\right)=(\mathrm{d}\theta)^{n}\,,
$$
where $dim(\mathbb{M})=n+1$.
Then, writing the contact structure $\theta$ as:
$$
\theta=dt-p_adq^a\,,
$$
the coordinate expression of the Jacobi bracket reads
$$
\left\lbrace f,g \right\rbrace_{(\Lambda,\Gamma)}=\dfrac{\partial f}{\partial p_a}\dfrac{\partial g}{\partial q^a}-\dfrac{\partial f}{\partial q^a}\dfrac{\partial g}{\partial p_a}+\dfrac{\partial f}{\partial t}\left( g-p_a\dfrac{\partial g}{\partial p_a} \right) -  \left( f-\dfrac{\partial f}{\partial p_a}p_a \right)\dfrac{\partial g}{\partial t}\,.
$$
In the case of contact manifolds, we can say that the Jacobi bracket is the counterpart of contact structures just as the Poisson bracket is the counterpart of the symplectic structure for symplectic manifolds.

The Jacobi bracket turns out to be relevant for the geometrical formulation of thermodynamics (\cite{bravetti_lopez-monsalvo_nettel-contact_symmetries_and_hamiltonian_thermodynamics}) and for the covariant description of relativistic point particles.
\end{example}

\subsection{Clifford Algebras}
Given a vector space V, one can consider its tensor algebra 
$$
T(V)= \oplus_{k=0}^{k=\infty} V^{\otimes k}
$$
If the vector space V is equipped with a bilinear symmetric form $g$, one can consider the two sided ideal $I_g(V)$ generated by elements of the form 
$$
v\otimes v -g(v,v)1
$$ 
The quotient space 
$$
\dfrac{T(V)}{I_g(V)}= Cl(V,g)
$$ 
is the Clifford algebra associated to the space $(V,g)$. The tensor product on $T(V)$ induces a product on $Cl(V,g)$ that we will call $\vee$-product. If the vector space has dimension $n$, its Clifford algebra has dimension $2^n$ which is the same dimension as the exterior algebras $\Lambda(V)$ and $\Lambda(V^{*})$. Therein it is possible to establish a one-to-one correspondence between these vector spaces (\cite{kahler-der_innere_differentialkalkul}).
 Exploiting this correspondence we realize a Clifford algebra on $\Lambda(V^*)$ and $\Lambda(V)$ by defining the following product
$$
\phi \vee \omega = \sum_{s}(-1)^{\frac{s(s-1)}{2}}g^{a_1b_1}\cdots g^{a_sb_s}\left(\gamma^s \left( i_{a_1}\cdots i_{a_s} \phi \right)\wedge \left( i_{b_1}\cdots i_{b_s} \omega \right)  \right)
$$      
where $\phi \in \Lambda^k(V^*)$ and $\omega \in \Lambda^p(V^*)$; $i_{a_1}=i_{e_{a_1}}$ is the contraction with $\left\lbrace e_a \right\rbrace$, defining an orthonormal basis of the vector space V. \newline
If V has an algebra structure as well, previous constructions can be extended to this setting and we can analyze the behaviour of the algebraic differential calculus with respect to a Clifford algebra structure. It is possible to show that $i_v$ is still a graded derivation of the $\vee$-product with degree $-1$. On the other side exterior derivative $d$ is not a derivation of this new product, whereas the Lie derivative $L_v$ is a derivation iff the vector $v$ is in the centre of the Lie algebra V. \newline
Let us consider an orthonormal basis $\left\lbrace e_a \right\rbrace$ of the vector space V; we will denote its dual basis by $\left\lbrace e^a \right\rbrace$. Then it is possible to define an antisymmetric tensor of maximal degree on V as 
$$
\omega=e^1\wedge \cdots \wedge e^n
$$  
which we will call a volume form. This choice allows us to define an operator $\ast :  \Lambda^k(V^*)\; \rightarrow \; \Lambda^{n-k}(V^*)$ which associates to the k-form $\alpha=\alpha_{a_1\cdots a_k}e^{a_1}\wedge \cdots \wedge e^{a_k}$, the $n-k$-form given by
$$
\ast \alpha (v_1,\cdots , v_{N-k})=\alpha_{a_1\cdots a_k}g^{a_1b_1}\cdots g^{a_kb_k}\omega(e_{b_1}\wedge \cdots \wedge e_{b_k}\wedge v_1 \wedge \cdots \wedge v_{n-k})
$$  
By using $\ast$-operator one can build another homology operator which decreases the degree of the antisymmetric tensor on which acts, say 
\begin{eqnarray*}
\delta : \Lambda^{k}(V^*)\rightarrow \Lambda^{k-1}(V^*)\\
\delta \alpha = (-1)^{n-k}(-1)^{sign(g)}\ast d \ast \alpha
\end{eqnarray*} 
Having extended our algebraic calculus, we are able to define a possible Dirac operator as 
$$
D=(d+\delta)
$$
It will act on the exterior algebra, which will play the role of the space of spinors\cite{graf-differential_forms_as_spinors}.

\subsection{Clifford Modules}
By using the same ideology of gluing together vector spaces at each point of a manifold $\mathbb{M}$, we construct the exterior algebra $\mathcal{A}(\mathbb{M})$ over $\mathbb{M}$ which is 
$$
\mathcal{A}(\mathbb{M})=\cup_m \Lambda(T^*_m\mathbb{M})
$$
Practically this means that, as already explained, the field $\mathbb{K}$ of coefficients of the vector space V is replaced by the ring of functions $\mathcal{F}(\mathbb{M})$, and the sections $\Lambda(\mathbb{M})=\Gamma(\mathcal{A}(\mathbb{M}))$ form a module on this ring. If the manifold is equipped with a bilinear symmetric form we can define a $\vee$-product at each point and the resulting bundle $\mathcal{A}(\mathbb{M})$ will have also a Clifford algebra representations. This means that $\Lambda(\mathbb{M})$ is a Clifford module (\cite{lawson_michelson-spin_geometry}). \newline
As before we can analyze the behaviour of the exterior differential calculus with respect to this Clifford structure. Once again the contraction $i_X$ remains a derivation of degree $-1$; the Lie derivative $L_X$ is a derivation if $X$ is a Killing vector field; the exterior derivative $d$ is not a derivation. \newline
\begin{remark}
If g is a nondegenerate symmetric bilinear form it is associated to a metric tensor, which we will call g. If one considers the Levi-Civita connection on the tangent bundle TM, a covariant derivative is defined. It can be shown that the covariant derivative $\nabla_X$ is a derivation of the $\vee$-product. As this product does not increase the support of the sections, this derivations are associated to vector fields, and this construction can be also described in the algebraic setting of Lie Algebroids (\cite{mackenzie-general_theory_of_lie_groupoids_and_algebroids}). 
\end{remark}
Let us come back to $\Lambda(\mathbb{M})$. Let us consider a bilinear form which is also non-degenerate. By exploiting the scalar product defined at each point, we can introduce on $\Lambda(\mathbb{M})$ a $\mathcal{F}(\mathbb{M})$-valued scalar operator
\begin{eqnarray*}
(\cdot , \cdot ): \Lambda^k(\mathbb{M})\times \Lambda^k(\mathbb{M}) \rightarrow \mathcal{F}(\mathbb{M})\\
(\alpha,\beta)= \ast (\alpha \wedge \ast \beta) 
\end{eqnarray*}
and saying that it is zero if the two forms have different degrees. $\ast$ is the Hodge dual operator which associates to each k-form $\alpha= \alpha_{a_1\cdots a_k}\theta^{a_1}\wedge \cdots \wedge \theta^{a_k}$, the $(n-k)$-form
$$
\ast \alpha = \epsilon^{a_1\cdots a_k}_{b_{k+1} \cdots b_n}\alpha_{a_1 \cdots a_k}\theta^{b_{k+1}}\wedge \cdots \wedge \theta^{b_n} \, .
$$ 
It is related to the choice of the volume form $\Omega$  
$$ 
\Omega=\theta^1\wedge\theta^2\wedge\cdots \wedge \theta^n \, ,
$$
where $\left\lbrace \theta^a \right\rbrace$ is an orthonormal basis of the module $\Lambda^1(M)$. \newline
If we want a scalar product we can define 
\begin{equation}
\langle\alpha,\beta\rangle=\int_M (\alpha,\beta) \Omega
\label{scal}
\end{equation}
The next step is the introduction of another boundary operator, say
\begin{eqnarray*}
\delta : \Lambda^{k}(M)\rightarrow \Lambda^{k-1}(M)\\
\delta \alpha = (-1)^{n-k}(-1)^{sign(g)}\ast d \ast \alpha
\end{eqnarray*}
In terms of $d$ and $\delta$, the Laplace-Beltrami operator $\Delta$ can be written as 
$$
\Delta= d\delta + \delta d=(d+\delta)^2
$$
Therefore a Dirac-like operator is given by the scalar differential operator 

\be\label{eqn: dirac-like operator}
D= (d+\delta)
\ee
as it is a square-root of the Laplacian.\newline
Let us note two things: 
\begin{itemize}
\item First of all one can notice that if $M$ has no boundary $\delta$ is the adjoint of $d$ with respect to the scalar product $\eqref{scal}$; on the contrary, if $M$ has a (smooth) boundary $\partial M$, then $\delta$ is not the adjoint of $d$ and the two operators are related by
$$
\langle \mathrm{d}\alpha\,,\beta\rangle - \langle \alpha\,,\delta\beta\rangle=\int_{M}\mathrm{d}\left(\alpha\wedge\star\beta\right)=\int_{\partial M}i^{*}\left(\alpha\wedge\star\beta\right)
$$
where $i\colon\partial M\rightarrow M$ denotes the immersion of $\partial M$ in $M$.
\item The same procedure can be applied also if g has a kernel. The main change appears in the definition of a suitable $\ast$-operator and consequently of a $\delta$.   
\end{itemize}

\begin{remark}
In the context of  differential equations written on the space of differential forms, an useful theorem is the so-called Hodge decomposition theorem \cite{carinena_ibort_marmo_morandi-geometry_from_dynamics_classical_and_quantum}. 
It states that the Hilbert space of differential forms can be decomposed into the direct sum of three orthogonal subspaces which are respectively:

\begin{itemize}
\item the space of exact differential forms, that is, forms $\alpha \in \Lambda^{k}(\mathbb{M})$ which can be written as $\mathrm{d}\beta$, whith $\beta \in \Lambda^{k-1}(\mathbb{M})$; 
\item the space of coexact forms, that is, forms $\alpha \in \Lambda^{k}(\mathbb{M})$ such that $\alpha = \delta \beta$ with $\beta \in \Lambda^{k+1}(\mathbb{M})$; 
\item the space of harmonic forms, i.e., differential forms $\alpha$ such that $\Delta \alpha = 0$.
\end{itemize} 
A consequence of this theorem is the well known fact that, on $\mathbb{R}^{3}$, ``every vector field $X$ can be written as the sum of a gradient of something plus the curl of something else''.
Here, a gradient vector field $Y$ is defined as $G(\mathrm{d}f\,,\cdot)$, where $f$ is a function and $G$ is the inverse of a metric tensor $g$; and a curl vector field $Z$ is defined as $G(\delta\beta\,,\cdot)$, where $\beta$ is a differential two-form.

\end{remark}

\begin{remark}
The expression in equation (\ref{eqn: dirac-like operator}) for a Dirac-like operator is scalar with respect to the diffeomorphisms group.
There are only other two derivations of the wedge product which are covariant with respect to this group, i.e., the interior product (or contraction) $i_{X}$, and the Lie derivative $\mathcal{L}_{X}$, where $X$ is a generic vector field.
Interior product is a derivation of degree $-1$, and its action is defined as follows

\bea
& i_{X}\colon \Lambda^{k}(\mathbb{M})\rightarrow\Lambda^{k-1}(\mathbb{M}) \\
& \alpha\mapsto i_{X}\alpha \\
& i_{X}\alpha(X_{1}\,,...,X_{k-1}):=\alpha(X\,,X_{1}\,,...\,,X_{k-1})\,.
\eea
The Lie derivative is a derivation of degree $0$, and its action on differential forms can be characterized by means of the Cartan formula:

\bea
& \mathcal{L}_{X}\colon \Lambda^{k}(\mathbb{M})\rightarrow\Lambda^{k}(\mathbb{M}) \\
& \mathcal{L}_{X}=i_{X}\,\mathrm{d} + \mathrm{d}\,i_{X}\,.
\eea
According to the scalar product on the space of differential forms, these operators are not symmetric.
It is well known (\cite{abraham_marsden_ratiu-manifolds_tensor_analysis_and_applications}) that the adjoint of $i_{X}$ is $i_{X}^{\dagger} = (-1)^{n(k+1)+1}\ast i_{X}\ast$, when it acts on a k-form. In particular, 

\be
(i_{X} \alpha , \beta) = (\alpha , i_{X}^{\dagger}\beta) = (\alpha , G(X, \cdot) \wedge \beta )
\ee
where $\alpha \in \Lambda^{1}(\mathbb{M})$, $\beta$ is a function, and $G$ is the inverse of the metric tensor $g$.

On the other side, the adjoint of $\mathcal{L}_{X}$ is $\mathcal{L}_{X}^{\dagger}= (-1)^{k(n-k)+1}\ast \mathcal{L}_{X} \ast$ and it satisfies the following relation 

\be
(\, \mathcal{L}_{X}\alpha \, , \beta \,) - (\,\alpha \, ,  \mathcal{L}_{X} ^{\dagger} \beta \,) = \int_{\partial \mathbb{M}} i_{X}(\alpha \wedge \ast \beta)\, ,
\ee
where the right-hand side appears only in the case of manifolds with a boundary.
In particular, when $\alpha , \beta $ are functions, the adjoint $\mathcal{L}_{X}^{\dagger}$ is the first order differential operator $\mathcal{L}_{X} + div(X)$, with $div(X)$ a scalar function defined by:

\be
\mathcal{L}_{X}\Omega = div(X) \,\Omega\,,
\ee
where $\Omega$ is the volume form associated with the metric tensor $g$.
Thus, on functions, we use only $\mathcal{L}_{X} + \frac{1}{2}\,div(X)$ to deal with a symmetric differential operator.

The three graded derivations $\mathrm{d}\,,i_{X}\,,\mathcal{L}_{X}$ close a graded Lie algebra according to the following rules:

\be
\left[i_{X}\,,\mathcal{L}_{Y}\right]= i_{[X\,,Y]}\,,\;\;\;\;\;\; \left[i_{X}\,,\mathrm{d}\right]=\mathcal{L}_{X}\,,\;\;\;\;\;\;\left[\mathrm{d}\,,\mathcal{L}_{X}\right]=0\,.
\ee

The study of these differential operators is fundamental when one deals with theories that are covariant with respect to the diffeomorphisms group, such as general relativity.

\end{remark}

\begin{example}
An interesting application of the concept of duality in order to define a differential calculus can be found in the discrete exterior differential calculus built by Desbrun et al in \cite{desbrun_hirani_leok_marsden-discrete_exterior_calculus}. Their idea is to define a differential calculus intrinsically discrete which mimics the usual differential calculus on continuous manifolds but it is not derived by means of approximation procedures. 
At this purpose let us start with a manifold made up of a set of points and consider the complex $\mathcal{K}$ (see for instance \cite{munkres-elements_of_algebraic_topology}) of dimension $n$ generated by linking them one to each other. Since we are interested in a Dirac-Kahler operator in this discrete setting, let us introduce a local metric on this complex by defining a distance map among vertices of each $n$-simplex in the complex. This local metric is a map $d\, \colon \, \left\lbrace (v_j,\, v_k) , \, \forall \left[ v_j v_k \right] \in \sigma^n \right\rbrace \, \rightarrow \, \mathbb{R} $ must satisfy some properties:
\begin{itemize}
\item Positive : $d(v_j,\, v_k) \geq 0, \,\, \mbox{ and } \, d(v_j, \, v_j) = 0$
\item Strictly Positive : $d(v_j, \, v_k )= 0 \, \Rightarrow \, v_j = v_k$
\item Simmetry : $d(v_j, \, v_k) = d(v_k , \, v_j)$
\item Triangle inequality : $d(v_j, \, v_l) \leq d(v_j , \, v_k) + d(v_k, \, v_l )$
\end{itemize}
In other words we are considering any single $n$-simplex as an open subset of the discrete manifold: as already said, indeed, every open subset of a manifold can be identified with its tangent space. Giving a local metric, therefore, is equivalent to the assignment of a scalar product on the tangent space at each point.    

Let us call $C_k(\mathcal{K};\mathbb{Z})$ the space of $k$-chains, that is the free abelian group of formal finite sums of $k$-simplices in $\mathcal{K}$ with coefficients in $\mathbb{Z}$. The space of Homomorphism from $C_k(\mathcal{K}; \mathbb{Z})$ to the additive group $\mathbb{R}$ is the space of cochains and will be denoted as $\Omega^k(\mathcal{K})$. In this discrete setting the space of cochains replaces the space of differential forms. The pairing
between a $k$-chain $\sigma^k$ and a $k$-cochains $\alpha^k$ will be denoted by 
$$
\alpha^k(\sigma^k) = \left\langle \alpha^k , \sigma^k \right\rangle \, .
$$
It can be thought as the result of integrating a differential forms along a path. 

A further ingredient is the so called dual complex, that is the complex which is made up of circumcentres (or barycentres) of each $n$-simplex of the complex. By means of this dual subdivision one can associate to each $k$-simplex a $(n-k)$-simplex of the dual complex 
\begin{equation}
\label{dual simplex}
\star(\sigma^k) = \sum_{\sigma^j, \, j\geq k} \varepsilon_{\sigma^k} \left[ c(\sigma^k) \cdots c(\sigma^j) \cdots c(\sigma^n))\right]
\end{equation} 
where $c(\sigma^j)$ is the circumcentre of the simplex $\sigma^j$ and $\left[ c(\sigma^k) \cdots c(\sigma^j) \cdots c(\sigma^n)) \right]$ denotes the convex hull of the points $c(\sigma^j)$ with $j\geq k$ and $\varepsilon_{\sigma^k}$ is a factor related with the orientation of the dual simplex. This dual complex will play a fundamental role in the definition of a Hodge operator.

Coming back to the space of cochains, it can be equipped with the wedge product $\wedge \, \colon \, \Omega_d^k(\mathcal{K}) \times \Omega_d^l(\mathcal{K}) \, \rightarrow \, \Omega_d^{k+l}(\mathcal{K})$ defined by

$$
\left\langle \alpha^k \wedge \beta^l , \sigma^{k+l} \right\rangle = \dfrac{1}{(k+l)!}\sum_{\tau \in S_{k+l+1}} \mathrm{sign}(\tau) \dfrac{|\sigma^{k+l} \cap \star v_{\tau(k)}|}{|\sigma^{k+l}|}
$$
\begin{equation}
\label{wedge}
\left\langle \alpha^k \, , \, \left[ v_{\tau(0)} \cdots v_{\tau(k)}\right]  \right\rangle  \left\langle \beta^l \, , \, \left[ v_{\tau(k+1)}\cdots v_{\tau (k+l)}  \right]  \right\rangle
\end{equation}
where $|\sigma^{k}|$ denotes the volume of the $k$-simplex calculated according to the local distance and the local embedding in $\mathbb{R}^n$. 

The boundary operator $\partial_k \, \colon \, C_k(\mathcal{K},\mathbb{Z}) , \rightarrow \, C_{k-1}(\mathcal{K},\mathbb{Z})$ is a homomorphism defined by its action on the simplex $\sigma^k = \left[ v_0,\cdots, v_k  \right]$ 
\begin{equation}
\label{boundary}
\partial_k \sigma^k = \sum_{j=0}^k (-1)^j \left[ v_0, \cdots , \hat{v}_j, \cdots v^k  \right] \, ,
\end{equation}
where $\left[ v_0, \cdots , \hat{v}_j, \cdots v^k  \right]$ is the simplex obtained by omitting the vertex $v_j$.

The exterior derivative $d_k \, \colon \, \Omega_d^k(\mathcal{K}) \, \rightarrow \, \Omega^{k+1}_d(\mathcal{K}) $ is defined as the coboundary operator
\begin{equation}
\label{coboundary}
\left\langle d_k \alpha^k\, , \,\sigma^{k+1}  \right\rangle = \left\langle \alpha^k\, , \,\partial_{k+1} \sigma^{k+1}  \right\rangle \, .
\end{equation}   
This expression makes Stoke's theorem true by definition. 

It is possible to show that exterior differential $d \, \colon \, \Omega_d(\mathcal{K}) \, \rightarrow \, \Omega_d (\mathcal{K})$ is a derivation of the wedge product but this product, in general, is no more associative; however if we restrict to the subset of closed cochains the product is associative. 

Since there exists a dual complex one can define the discrete Hodge Star operator as the map $\ast \, \colon \, \Omega_d^k(\mathcal{K}) \, \rightarrow  \, \Omega^{n-k}_d(\mathcal{K})$ which acts as follows
\begin{equation}
\label{Hodge}
\dfrac{1}{|\sigma^k|} \left\langle \alpha^k \, , \, \sigma^k \right\rangle = \dfrac{1}{|\star \sigma^k |}\left\langle \ast \alpha \, , \, \star \sigma^k \right\rangle
\end{equation}

Now the codifferential operator is $\delta \, \colon \, \Omega_d^k(\mathcal{K}) \rightarrow \Omega_d^{k-1}(\mathcal{K})$ 
\begin{equation}
\label{codifferential}
\delta \beta^k = (-1)^{nk+1} \ast d \ast \beta^k
\end{equation} 

Having at our disposal all the necessary ingredients, we can explicitly write and try to solve Dirac-Kahler equation for a one-dimensional discrete manifold.  
In particular let us consider a complex made up of the union of an infinite number of one-dimensional simplices $\left[ v_j , v_{j+1} \right]$. For simplicity, let us consider a local metric which is the same at each point, that is $d(v_j, v_{j+1}) = d(v_k, v_{k+1}) = l$. If we denote the dual elements of the simplices $v_j$ and $\sigma_j=\left[ v_j v_{j+1}\right]$ by $\mathbf{e}_j^{(0)}$ and $\mathbf{e}_j^{(1)}$, respectively, one can write $0$-cochains and $1$-cochains as
\begin{eqnarray*}
&\alpha^{(0)}= \sum_j \alpha_j^{(0)} \mathbf{e}_j^{(0)} \\ 
&\alpha^{(1)}= \sum_j \alpha_j^{(1)} \mathbf{e}_j^{(1)} \,  .
\end{eqnarray*}
Therefore one can write Dirac-Kahler operator acting on a generic cochain as:
\begin{equation}
\label{discreteDK}
\mathrm{i}(d-\delta) (\,\alpha^{(0)} + \alpha^{(1)}\,)\, = \mathrm{i}\, \sum_j \left((\alpha^{(1)}_{j}-\alpha^{(1)}_{j-1})\mathbf{e}_j^{(0)} + (\alpha^{(0)}_{j+1}-\alpha^{(0)}_j)\mathbf{e}_j^{(1)} \right)
\end{equation}
We can diagonalize this operator by choosing 
\begin{eqnarray*}
&\alpha_j^{(0)}=A \mathrm{e}^{-\mathrm{i}jlk} \\
&\alpha_j^{(0)}=B \mathrm{e}^{-\mathrm{i}jlk}
\end{eqnarray*}
If one replaces this ansatz in the eigenvalue equation 
\begin{equation*}
\mathrm{i}(d-\delta) (\,\alpha^{(0)} + \alpha^{(1)}\,) = \lambda (\,\alpha^{(0)} + \alpha^{(1)}\,)
\end{equation*}
one gets the following pair of equation for the coefficients $A$ and $B$:

\begin{eqnarray*}
&\mathrm{i}A (1-\mathrm{e}^{\mathrm{i}kl})-\lambda B = 0 \\
&-\lambda A+ \mathrm{i}B (\mathrm{e}^{-\mathrm{i}kl}-1) = 0
\end{eqnarray*}

This system admits a non vanishing solution if and only if the determinant of the coefficient matrix is zero, that is
$$
\lambda ^2 + \mathrm{e}^{-\mathrm{i}kl}\left( 1 - \mathrm{e}^{\mathrm{i}kl} \right)^2 = 0 \, ,
$$
whose solutions are 
$$
\lambda = \pm \sqrt{2\left(1-\cos (kl)\right)} = \pm 2 | \sin \left( \frac{kl}{2} \right) |
$$

This simple expressions, which agrees with the results provided by other models describing fermions on a lattice, has been obtained by considering a constant metric. However one can generalize the expression \eqref{discreteDK} to the case of non constant metric and eventually we get the result that local metric behaves like a coupling constant for this system.
It is possible to use this calculus to deal with gauge theories on lattices.
However, the study of such a model needs a deeper investigation, which falls outside the purpose of this paper and will be presented somewhere else.   

\end{example}

\section{Manifolds with boundaries and corners} \label{section: space of states}

In accordance with the $C^{*}$-algebraic approach to classical and quantum theories, the manifolds we shall be interested in arise as convex subsets of linear spaces.
For simplicity, we shall consider finite-dimensional systems.
Specifically, the states of classical or quantum systems in the $C^{*}$-algebraic approach are the normalized, positive, real linear functionals on the $C^{*}$-algebra $\calg$ of the system.

In the finite-dimensional classical case, that is, when $\calg$ is commutative with dimension $N<+\infty$, the states of the system can be identified with the probability vectors, namely, the vectors $\vec{p}\equiv(p_{1},p_{2},...,p_{N})$ in $\mathbb{R}^{N}$ with the properties $p_{j}\geq0$ and $\sum_{j}p_{j}=1$.
Clearly they form a manifold with corners arising from the Cartesian product of $N$ copies of $\mathbb{R}_{+}$ intersected with the affine space $\sum_{j}p_{j}=1$.

There is a simple way to ``quantize'' a classical probability vector, indeed, let us recall that if $|\psi\rangle\in\mathbb{C}^{N}$ and $\sum_{j}|e_{j}\rangle\langle e_{j}|=\mathbb{I}$ is an orthonormal resolution of the identity, we have a probability vector associated with any $|\psi\rangle$ and any partition by setting:

\be
p_{j}:=\frac{tr\left(|\psi\rangle\langle\psi|e_{j}\rangle\langle e_{j}|\right)}{\langle\psi|\psi\rangle}=\frac{\langle\psi|e_{j}\rangle\langle e_{j}|\psi\rangle}{\langle\psi|\psi\rangle}\,.
\ee
The Euclidean structure of $\mathbb{C}^{N}$ defines a distance on the space of probability vectors which is associated with the metric:

\be
ds^{2}=\sum_{j,k}g_{jk}\,dp^{j}dp^{k}=\frac{1}{4}\sum_{j=1}^{N+1} \frac{dp^{j}dp^{k}}{p^{j}}\delta_{jk}\,.
\ee
This is the so-called Fisher-Rao metric.
We notice that if we set $x^{j}:=\sqrt{p^{j}}$, we get $1=\sum_{j}p_{j}=\sum_{j}x_{1}^{2} + x_{2}^{2}+\cdots+x_{N}^{2}$, hence, we would get the positive hyperoctant of $S^{N-1}$.
For example, the flat simplex for $N=3$ becomes the round octant of the sphere.
Furthermore, the differential of the $x^{j}$ are $dx^{j}=\frac{dp^{j}}{2\sqrt{p^{j}}}$, and thus, the Fisher-Rao metric in appropriate coordinates reads $ds^{2}=\sum_{j}dx^{j}dx^{j}$ (see \cite{bengtsson_zyczkowski-geometry_of_quantum_states:_an_introduction_to_quantum_entanglement}).

\begin{remark}
In setting $x^{j}:=\sqrt{p^{j}}$ we have decided to consider the square root within the real numbers.
However, if we consider $z^{j}=\mathrm{e}^{\imath\varphi_{j}}\sqrt{p^{j}}$ we would go back to the quantum picture where $z^{j}=\langle e_{j}|\psi\rangle$, i.e., we would reconstruct a vector in $\mathbb{C}^{N+1}$ and an Hermitean tensor:


$$
\mathbf{h}=\frac{1}{4}\left[\langle \mathrm{d}(\ln \vec{p})\otimes \mathrm{d}(\ln \vec{p})\rangle_{\vec{p}} - \langle \mathrm{d}(\ln \vec{p})\rangle_{\vec{p}}\otimes \langle\mathrm{d}(\ln \vec{p})\rangle_{\vec{p}} \right] +
$$
\be
+ \langle \mathrm{d}\vec{\varphi} \otimes\mathrm{d}\vec{\varphi}\rangle_{\vec{p}} - \langle \mathrm{d}\vec{\varphi}\rangle \otimes \langle \mathrm{d}\vec{\varphi}\rangle_{\vec{p}} + \frac{\mathrm{i}}{2}\left[\langle \mathrm{d} \left(\ln \vec{p}\right)\wedge\mathrm{d}\vec{\varphi}\rangle_{\vec{p}} - \langle \mathrm{d}\left(\ln \vec{p}\right)\rangle_{\vec{p}}\wedge\langle\mathrm{d}\vec{\varphi}\rangle_{\vec{p}}\right]\,,
\ee
where $\langle\,\rangle_{\vec{p}}$ denotes the expectation value with respect to the probability vector $\vec{p}$, and $\langle \mathrm{d}\ln\vec{p}\rangle$ is actually zero because of the constraint $\sum_{j}p_{j}=1$.
This is the pullback to the Hilbert space $\mathcal{H}$ of the canonical Hermitean tensor on the complex projective space $\mathcal{P}(\mathcal{H})$ (\cite{facchi_kulkarni_manko_marmo_sudarshan_ventriglia-classical_and_quantum_fisher_information_in_the_geometrical_formulation_of_quantum_mechanics}).
\end{remark}

A different way to ``quantize'' classical probability vectors would be to consider a probability vector as a diagonal matrix in the $C^{*}$-algebra $\calg=\mathcal{B}(\mathcal{H})$ and act on it with elements of the unitary group $U(N)$:

\be
\rho=\mathbf{U}^{\dagger}\,\left(\begin{array}{ccc}p_{1} & & \\ & \ddots \\ & & p_{N}\end{array}\right)\,\mathbf{U}\,,
\ee
thus obtaining a ``polar'' representation of a generic mixed state associated with $\mathbb{C}^{N}$.
In this picture, states becomes normalized, positive Hermitean matrices acting on $\mathbb{C}^{N}$.
The space $\stsp$ of quantum states is not a differential manifold, nor a differential manifold with smooth boundary.
Specifically, $\stsp$ is a subset of the dual $\mathfrak{u}^{*}(N)$ of the Lie algebra $\mathfrak{u}(N)$ of the unitary group $U(N)$, and it can be shown (\cite{grabowski_kus_marmo-geometry_of_quantum_systems_density_states_and_entanglement, grabowski_kus_marmo-symmetries_group_actions_and_entanglement}) that it is the union of different strata $\stsp^{k}$ labelled by the rank of the quantum states they contain. 
At the ``classical'' level, all the probability vectors with the same number of non-zero elements  would be on a given face\footnote{Recall that a face of a convex set $K$ is a convex subset $F\subset K$ such that, given $x,y\in K$, if $\alpha x + (1-\alpha)y\in F$ then $x,y\in F$.} (corner) if they occupy the same position in the column matrix describing a state.
At the quantum level, probability vectors with the same elements can be mapped one into the other by means of the unitary group, and if we allow for a nonlinear action of the complexification of the unitary group, we can map any state of a given rank into any other state of the same rank.
Thus, each stratum is an orbit, with respect to a non-linear action, of the complexification of $SU(N)$. 
This shift from $U(N)$ to $SU(N)$ comes from the coniugacy action on states.

In what follows we are going to consider the geometry of $\stsp$ according to \cite{grabowski_kus_marmo-geometry_of_quantum_systems_density_states_and_entanglement, grabowski_kus_marmo-symmetries_group_actions_and_entanglement, grabowski_kus_marmo-on_the_relation_between_states_and_maps_in_infinite_dimensions,  bengtsson_zyczkowski-geometry_of_quantum_states:_an_introduction_to_quantum_entanglement}.
Then, considering the geometrization of the dynamics of open quantum systems, we will argue that a notion of vector fields (derivations) transversal to the strata is needed in order to describe physically interesting processes such as decoherence, for which the rank of the states is not preserved.

\subsection{The geometry of $\obsp^{*}$ and $\stsp$}

Our stratified manifold, the space of quantum states $\stsp$, is a convex subset of the set $\obsp^{*}$ which is the dual space of the real vector space $\obsp$ defined by the real elements of a $C^{*}$-algebra with unity $\calg$.
The elements of $\obsp$ are identified with the observable of the theory.
We shall consider only the case in which $\calg$ is a finite-dimensional simple $C^{*}$-algebra with unity, in which case it is necessarily isomorphic to the set $\mathcal{B}(\mathcal{H})$ of bounded linear operators on a complex Hilbert space $\mathcal{H}$ of dimension $\left(dim(\mathcal{H})\right)^{2}=dim(\calg)=N$ (see \cite{davidson-c*_algebras_by_example}).
In this case, $\obsp$ becomes isomorphic to the set of self-adjoint linear operators on $\mathcal{H}$.
We shall limit ourselves to a quick revision of the geometry of $\obsp^{*}$ and $\stsp$, we refer to \cite{grabowski_kus_marmo-geometry_of_quantum_systems_density_states_and_entanglement, grabowski_kus_marmo-symmetries_group_actions_and_entanglement, grabowski_kus_marmo-on_the_relation_between_states_and_maps_in_infinite_dimensions,  bengtsson_zyczkowski-geometry_of_quantum_states:_an_introduction_to_quantum_entanglement} for a detailed discussion.

As said before, the elements of the linear space $\obsp$ are identified with observables, and the space $\stsp$ of quantum states is a convex subset of the dual space $\obsp^{*}$ of the space of observables.
Specifically, denoting with $\mathbb{I}$ the identity element in $\calg$, an element $\rho\in\obsp^{*}$ is a quantum states if $\rho(\mathbf{a})\geq0$ for every $\mathbf{a}\in\obsp$ with positive semi-definite spectrum\footnote{This condition is equivalent to $\mathbf{a}=\mathbf{b}^{2}$ for some $\mathbf{b}\in\obsp$, or, when $\mathbf{a}$ is looked at as an element of $\calg$, it is equivalent to the existence of $\mathbf{b}\in\calg$ such that $\mathbf{a}=\mathbf{b}\mathbf{b}^{\dagger}$.}, and if $\rho(\mathbb{I})=1$.
We will describe the geometry of $\stsp$ exploiting the ambient space $\obsp^{*}$, which is a linear space, and the algebraic properties of the space of observables $\obsp$ of which the ambient space $\obsp^{*}$ is the dual space.

The linear space $\obsp$ is not an associative subalgebra of the $C^{*}$-algebra $\calg$.
However, it carries a Lie algebra structure:

\be
\left[\left[\mathbf{a}\,,\mathbf{b}\right]\right]=-\frac{\imath}{2}(\mathbf{ab} - \mathbf{ba})\;\,\mbox{ with }\;\,\mathbf{a},\mathbf{b}\in\obsp\subset\calg\,,
\ee
and a Jordan algebra structure:

\be
\mathbf{a}\odot\mathbf{b}=\frac{1}{2}(\mathbf{ab} + \mathbf{ba})\;\;\;\;(\mathbf{a}\odot\mathbf{b})\odot \mathbf{a}^{2}=\mathbf{a}\odot(\mathbf{b}\odot\mathbf{a}^{2})\;\,\mbox{ with }\;\,\mathbf{a},\mathbf{b}\in\obsp\subset\calg\,,
\ee
making up what is known as a Lie-Jordan algebra structure on $\obsp$ (\cite{Emch-mathematical_and_conceptual_foundations_of_20th_century_physics, landsman-between_classical_and_quantum}).
This means that $\obsp$ is an algebra with respect to both the products, and it is such that $[[\,,]]$ is a derivation for $\odot$:

\be
\left[\left[\mathbf{a}\odot\mathbf{b}\,,\mathbf{c}\right]\right]=\mathbf{a}\odot\left[\left[\mathbf{b}\,,\mathbf{c}\right]\right] + \left[\left[\mathbf{a}\,,\mathbf{c}\right]\right]\odot\mathbf{b}\,,
\ee
and it is such that the deviation from associativity of the Jordan product is proportional to the deviation from associativity of the Lie product:

\be\label{equation: associator of Jordan and Lie product}
\mathbf{a}\odot(\mathbf{b}\odot\mathbf{c}) - (\mathbf{a}\odot\mathbf{b})\odot\mathbf{c}=\lambda^{2}\left(\left[\left[\mathbf{a},\left[\left[\mathbf{b},\mathbf{c}\right]\right]\,\right]\right] - \left[\left[\,\left[\left[\mathbf{a},\mathbf{b}\right]\right],\mathbf{c}\right]\right]\right)\;\,\mbox{ with }\;\,\mathbf{a},\mathbf{b},\mathbf{c}\in\obsp\subset\calg\,.
\ee
The constant $\lambda$ is there to remind us that if we want to recover the associative product of the $C^{*}$-algebra $\calg$ we have to fine-tune the Jordan and Lie product, specifically, the associative product is recovered by setting $\mathbf{ab}=\mathbf{a}\odot\mathbf{b} +\imath\left[\left[\mathbf{a},\mathbf{b}\right]\right]$.
When the appropriate physical units are introduced, it turns out that $\lambda$ is proportional to Planck constant $\hbar$.

We will now represent $\obsp$ and its Lie-Jordan structure in the space $\mathcal{F}(\obsp^{*})$ of functions on $\obsp^{*}$.
This is possible because, given any element  $\mathbf{a}\in\obsp$, we can define a linear function $f_{\mathbf{a}}:\obsp^{*}\rightarrow\mathbb{R}$ given by $f_{\mathbf{a}}(\xi):=\xi(\mathbf{a})$ with $\xi\in\obsp^{*}$.
This means that, associated with a basis $\{\mathbf{a}_{j}\}$ in $\obsp$ there is the coordinate system $\{x_{j}\}$ on $\obsp^{*}$ given by $x_{j}(\xi):=f_{\mathbf{a}_{j}}(\xi)$, and, furthermore, being $\obsp$ and $\obsp^{*}$ finite-dimensional vector spaces, the differentials $\mathrm{d}x_{j}$ of the coordinate functions form a basis of the cotangent space $T^{*}_{\xi}\obsp^{*}$ at each point $\xi\in\obsp^{*}$.

It is immediate to see that $f_{\mathbf{a}+\mathbf{b}}=f_{\mathbf{a}} + f_{\mathbf{b}}$, and thus the linear space $\obsp$ is represented as a linear subspace $\mathcal{F}_{l}(\obsp^{*})$ of the linear space $\mathcal{F}(\obsp^{*})$.
The linear space $\mathcal{F}(\obsp^{*})$ of smooth functions on $\obsp^{*}$ has a natural structure of algebra $\left(\mathcal{F}(\obsp^{*})\,;+\,;\cdot\right)$ where the product $\cdot$ is the pointwise product $(f\cdot g)(\xi):=f(\xi)g(\xi)$, with $f,g\in\mathcal{F}(\obsp^{*})$ and $\xi\in\obsp^{*}$.
However, $\mathcal{F}_{l}(\obsp^{*})$ is not closed with respect to the pointwise product in $\mathcal{F}(\obsp)$, indeed, the product of two linear functions is a quadratic function.
Therefore, we have to define two new products on $\mathcal{F}_{l}(\obsp^{*})$ so that it becomes a Lie-Jordan algebra.
This can be done defining the following products:

\be
\left(f_{\mathbf{a}}\ast f_{\mathbf{b}}\right)(\xi)=f_{\left[\left[\mathbf{\mathbf{a}}\,;\mathbf{b}\right]\right]}(\xi)\,,
\ee 

\be
\left(f_{\mathbf{a}}\odot f_{\mathbf{b}}\right)(\xi)=f_{\mathbf{a}\odot \mathbf{b}}(\xi)\,.
\ee
As it is clear, $\ast$ represents the Lie product of $\obsp$, and $\odot$ the Jordan product, and it is a matter of straightforward calculations to show that, together, they make $\mathcal{F}_{l}(\obsp^{*})$ into a Lie-Jordan algebra.
This means that the unary operation $L_{f_{\mathbf{a}}}(f_{\mathbf{b}}):=f_{\mathbf{a}}\ast f_{\mathbf{b}}$ is a derivation of $\odot$ for all $f_{\mathbf{a}}\,,f_{\mathbf{b}}\in\mathcal{F}_{l}(\obsp^{*})$.

Now, since linear functions generate the cotangent space $T^{*}_{\xi}\obsp^{*}$ at each $\xi\in\obsp^{*}$, we can use $\ast$ and $\odot$ to define two contravariant tensor fields on $\obsp^{*}$:

\be
\Lambda(df_{\mathbf{a}}\,;df_{\mathbf{b}})(\xi):=\left(f_{\mathbf{a}}\ast f_{\mathbf{b}}\right)(\xi)=f_{\left[\left[\mathbf{a},\mathbf{b}\right]\right]}(\xi)\,,
\ee

\be
\mathcal{R}(df_{\mathbf{a}}\,;df_{\mathbf{b}})(\xi):=\left(f_{\mathbf{a}}\odot f_{\mathbf{b}}\right)(\xi)=f_{\mathbf{a}\odot\mathbf{b}}(\xi)\,.
\ee
Note that $\Lambda$ and $\mathcal{R}$ are tensor fields on the whole $\obsp^{*}$.
Selecting the coordinate system $\{x_{j}\}$ on $\obsp^{*}$ given by $x_{j}(\xi):=f_{\mathbf{a}_{j}}(\xi)$, where  $\{\mathbf{a}_{j}\}$ is a basis in $\obsp$, we obtain the coordinates expression of $\Lambda$ and $\mathcal{R}$:

\be
\Lambda=c_{jk}^{l}x_{l}\frac{\partial}{\partial x_{j}}\wedge\frac{\partial}{\partial x}\;\; \mbox{ and } \;\;\; \mathcal{R}=d_{jk}^{l}x_{l}\frac{\partial}{\partial x}\otimes\frac{\partial}{\partial x}\,.
\ee

These two tensor fields allow to define the notions of Hamiltonian and gradient vector fields on $\obsp^{*}$.
Specifically, given $f\in\mathcal{F}(\obsp^{*})$, we define the Hamiltonian vector field $\widetilde{X}_{f}$ and the gradient vector field $\widetilde{Y}_{f}$ as follows:

\be
\widetilde{X}_{f}:=\Lambda\left(\mathrm{d}f\,,\cdot\right) \;\; \mbox{ and } \;\;\;  \widetilde{Y}_{f}:=\frac{1}{\lambda}\mathcal{R}\left(\mathrm{d}f\,,\cdot\right)\,,
\ee
where $\lambda$ is the parameter appearing in formula (\ref{equation: associator of Jordan and Lie product}).
A direct calculation shows that, when $f,g$ are linear functions say $f_{a},f_{b}$, Hamiltonian and gradient vector fields satisfy the following commutation relations::

\be
\left[\widetilde{Y}_{f_{\mathbf{a}}}\,;\widetilde{Y}_{f_{\mathbf{b}}}\right]=-\widetilde{X}_{\{f_{\mathbf{a}}\,;f_{\mathbf{b}}\}} \,,
\ee
\be
\left[\widetilde{X}_{f_{\mathbf{a}}}\,;\widetilde{X}_{f_{\mathbf{b}}}\right]=\widetilde{X}_{\{f_{\mathbf{a}}\,;f_{\mathbf{b}}\}} \,,
\ee
\be
\left[\widetilde{X}_{f_{\mathbf{a}}}\,;\widetilde{Y}_{f_{\mathbf{b}}}\right]= \widetilde{Y}_{\{f_{\mathbf{a}}\,;f_{\mathbf{b}}\}}\,.
\ee
Consequently, being $\{f_{\mathbf{a}}\,;f_{\mathbf{b}}\}=f_{\mathbf{a}}\ast f_{\mathbf{b}}=f_{\left[\left[\mathbf{a}\,;\mathbf{b}\right]\right]}$ a linear function, the generalized distribution:

\be
\mathcal{D}_{gl}:=\left\{Z\in\mathfrak{X}(\obsp^{*})\,:\;\,Z=\widetilde{X}_{f_{\mathbf{a}}} \mbox{ or } Z=\widetilde{Y}_{f_{\mathbf{a}}}\;\mbox{ with } \,f_{\mathbf{a}}\in\mathcal{F}_{l}(\obsp^{*})\right\}
\ee
is integrable.

Being $\calg\cong\mathcal{B}(\mathcal{H})$, the space of observables $\obsp$ is the space of Hermitean matrices on $\mathcal{H}$, furthermore, being in finite-dimension, every choice of basis in $\obsp$ determines a non-canonical isomorphism $\obsp^{*}\cong\obsp$.
This means that we can identify an element $\xi\in\obsp^{*}$ with an Hermitean matrix $\widetilde{\xi}\in\obsp$.
With the help of this isomorphism and the work \cite{grabowski_kus_marmo-geometry_of_quantum_systems_density_states_and_entanglement}, we see that the leaves of the foliation generated by $\mathcal{D}_{gl}$ are the orbits of the $\mathcal{GL}(\mathcal{H})$ action on $\obsp^{*}$ induced by the $\mathcal{GL}(\mathcal{H})$ action $\widetilde{\xi}\mapsto \mathbf{T}\widetilde{\xi} \mathbf{T}^{\dagger}$ on $\obsp$.
Note that the distribution $\mathcal{D}_{\Lambda}$ consisting of Hamiltonian vector fields associated to linear functions is itself integrable, and the leaves of the associated foliation are the orbit of the $\mathcal{U}(\mathcal{H})$ action induced by $\widetilde{\xi}\mapsto \mathbf{U}\widetilde{\xi} \mathbf{U}^{\dagger}$.

The leaves of $\mathcal{F}_{gl}$ have different dimensions.
Define the rank $rank(\xi)$ of $\xi\in\obsp^{*}$ as the matrix rank of its representative $\widetilde{\xi}\in\obsp$.
If $rank(\xi)=k$, we write $k_{+}$ for the number of positive eigenvalues of $\widetilde{\xi}$, and $k_{-}$ for the number of negative eigenvalues, both counted with multiplicities.
Clearly, $rank(\xi)=k_{+} + k_{-}$.
The main result in \cite{grabowski_kus_marmo-geometry_of_quantum_systems_density_states_and_entanglement} is the following:

\begin{theorem} \label{thm: structure of the orbits of the GL(H) action}
The family:

\be
\obsp^{*}(k_{+},k_{-}):=\left\{\xi\in\obsp^{*}\;:\;\;k_{+},k_{-}\geq0\;,\;k=k_{+}+k_{-}\leq n\right\}
\ee
of subsets of $\obsp^{*}$ is exactly the family of orbits of the smooth $\mathcal{GL}(\mathcal{H})$ action on $\obsp^{*}$ introduced above.
Specifically, every $\obsp^{*}(k_{+},k_{-})$ is a smooth, connected submanifold of $\obsp^{*}$.
The tangent space $T_{\xi}\obsp^{*}(k_{+},k_{-})$ at $\xi$ is identified with the linear subspace of $\obsp$ the elements $B$ of which satisfy the condition:

\be
\langle Bx|y\rangle=0 \;\;\forall x,y\in Ker(\,\widetilde{\xi}\,)\,.
\ee 
\end{theorem}

\vsp

With the aid of theorem \ref{thm: structure of the orbits of the GL(H) action} it is possible to show that $\stsp$  is the disjoint union of smooth manifolds $\stsp^{k}\subset\obsp^{*}$. 

At this purpouse, let use denote with $\mathcal{P}_{\calg}$ the set of $\xi\in\obsp^{*}$ such that  $\xi(\mathbf{a}^{2})\geq0$ for $\mathbf{a}\in\obsp$.
This set is the cone of positive hermitean functionals on $\calg$.
The representative $\widetilde{\xi}\in\obsp$ of $\xi\in\mathcal{P}_{\calg}$ through the isomorphism $\obsp^{*}\cong\obsp$ introduced above is a positive semi-defined Hermitean operator on $\mathcal{H}$, hence, $\widetilde{\xi}=\mathbf{A}\mathbf{A}^{\dagger}$ for some $\mathbf{A}\in\mathcal{B}(\mathcal{H})$.
The set of all such operators is denoted as $\mathcal{P}(\mathcal{H})$.
Accordingly, it is clear that the $\mathcal{GL}(\mathcal{H})$ action introduced above preserves $\mathcal{P}_{\calg}$ since:

$$
\mathbf{T}\,\widetilde{\xi}\,\mathbf{T}^{\dagger}=\mathbf{T}\,\mathbf{A}\mathbf{A}^{\dagger}\,\mathbf{T}^{\dagger}=\mathbf{T}\mathbf{A}\,\left(\mathbf{T}\mathbf{A}\right)^{\dagger}\,.
$$
Furthermore, this action preserves the rank and the number of positive and negative eigenvalues of $\widetilde{\xi}$, that is, the rank of $\xi\in\mathcal{P}_{\calg}$.
Denoting with $\mathcal{P}_{\calg}^{k}$ the set of elements of $\mathcal{P}_{\calg}$ of rank $k$, theorem \ref{thm: structure of the orbits of the GL(H) action} applies, and thus $\mathcal{P}_{\calg}^{k}$ is a smooth, connected submanifold of $\obsp^{*}$, specifically, it is $\obsp^{*}(k,0)$, and its dimension is $(2nk - k^{2})$ with $n=dim(\mathcal{H})$.
The set $\mathcal{P}_{\calg}$ is the disjoint union of the smooth manifolds $\mathcal{P}_{\calg}^{k}$ for $0\leq k\leq n^{2}$.
Note that the top stratum $\mathcal{P}_{\calg}^{n}$, with $n=dim(\mathcal{H})$, is an open submanifold of $\obsp^{*}$.
Furthermore, an important result in \cite{grabowski_kus_marmo-geometry_of_quantum_systems_density_states_and_entanglement} states that, for every smooth curve $\gamma:\mathbb{R}\rightarrow\obsp^{*}$ such that $\gamma(\tau)\in\mathcal{P}_{\calg}$ for all $\tau$, $\gamma(\tau)\in\mathcal{P}_{\calg}^{k}$ implies $\dot{\gamma}(\tau)\in T_{\gamma(\tau)}\mathcal{P}_{\calg}^{k}$.

Recall that a quantum state $\rho$ is an element in $\mathcal{P}_{\calg}$ such that $f_{\mathbb{I}}(\rho)=1$.
The function $f_{\mathbb{I}}$ is regular on each $\mathcal{P}_{\calg}^{k}$, hence, $\stsp^{k}$ is a regular submanifold of $\mathcal{P}_{\calg}^{k}$.
Topologically, $\stsp^{k}$ is connected, and it holds $\mathcal{P}_{\calg}^{k}\cong\stsp^{k}\times\mathbb{R}^{+}$.
The dimension of $\stsp^{k}$ is $(2nk - k^{2} -1)$, and the tangent space $T_{\rho}\stsp^{k}$ at $\rho$ is identified with the linear subspace of $\mathfrak{u}^{*}(\mathcal{H})$ consisting of traceless elements $B$ satisfying the condition:

\be
\langle Bx|y\rangle=0 \;\;\forall x,y\in Ker(\,\widetilde{\rho}\,)\,.
\ee 
Again, the tangent vector $\dot{\gamma}(\tau)$ of every smooth curve $\gamma:\mathbb{R}\rightarrow\obsp^{*}$ lying entirely in $\stsp$ is tangent to the stratum $\stsp^{k}$ to which $\gamma(\tau)$ actually belongs.
This results will bring important physical consequences when dealing with the dynamical evolution of a quantum system generated by a Kossakowski-Lindblad operator (\cite{gorini_kossakowski_sudarshan-completely_positive_dynamical_semigroups_of_N-level_systems, lindblad-on_the_generators_of_quantum_dynamical_semigroups}).

\vsp

A direct calculation shows that the Hamiltonian vector field $\widetilde{X}_{f}$ associated with the function $f$  by means of $\Lambda$ is tangent to the affine subspace $f_{\mathbb{I}}(\xi)=1$ for all functions $f$. 
However, gradient vector fields do not preserve the affine subspace $f_{\mathbb{I}}(\xi)=1$.
Now we want to find contravariant tensor fields $\Lambda_{\stsp}$ and $\mathcal{R}_{\stsp}$ such that the Hamiltonian and gradient vector fields associated with them are tangent to $\stsp^{k}$ for all $k$.
A natural way to proceed would be to use Dirac's prescription for Poisson brackets with constraints, however, we will follow a different path.
At this purpouse, we start considering the so-called expectation value functions $e_{\mathbf{a}}(\xi)=\frac{f_{\mathbf{a}}(\xi)}{f_{\mathbb{I}}(\xi)}$ on it.
Note that these functions are not defined on the whole $\obsp^{*}$ but, in general, only on the open submanifold $\mathcal{O}$ of $\obsp^{*}$ composed by elements $\xi\in\obsp$ such that $\xi(\mathbb{I})\neq0$.
We choose the expectation value functions because these functions are invariant with respect to dilations, and the affine subspace $f_{\mathbb{I}}(\xi)=1$ can be identified with the quotient of $\mathcal{O}$ with respect to the action of the dilations.
However, since $e_{\mathbb{I}}=1$, the expectation value functions are not enough to generate the cotangent bundle of $\mathcal{O}$, therefore, in order to correctly define tensors on $\mathcal{O}$, we have to consider the function $f_{\mathbb{I}}$.

Now, we define the tensor fields $\Lambda_{\stsp}$ and $\mathcal{R}_{\stsp}$ requiring that:

\be
\begin{array}{l}\Lambda_{\stsp}(\mathrm{d}e_{\mathbf{a}},\mathrm{d}e_{\mathbf{b}})(\rho)=e_{[[\mathbf{a},\mathbf{b}]]}(\rho)\,,\\ \\ \Lambda_{\stsp}(\mathrm{d}e_{\mathbf{a}},\mathrm{d}f_{\mathbb{I}})(\rho)=0\;\forall \mathbf{a}\in\obsp\,,\end{array}
\ee

\be
\begin{array}{l}\mathcal{R}_{\stsp}(de_{\mathbf{a}},de_{\mathbf{b}})(\rho)=e_{\mathbf{a}\odot \mathbf{b}}(\rho) - e_{\mathbf{a}}(\rho)e_{\mathbf{b}}(\rho)\,,\;\forall \mathbf{a},\mathbf{b}\neq\mathbb{I}\,, \\ \\ \mathcal{R}_{\stsp}(\mathrm{d}e_{\mathbf{a}},\mathrm{d}f_{\mathbb{I}})(\rho)=\mathcal{R}_{\stsp}(\mathrm{d}f_{\mathbb{I}}\,,\mathrm{d}e_{\mathbf{a}})(\rho)=\mathcal{R}_{\stsp}(\mathrm{d}f_{\mathbb{I}}\,,\mathrm{d}f_{\mathbb{I}})(\rho)=0\;\forall \mathbf{a}\in\obsp\,.\end{array}
\ee
Note that the Jordan algebra structure of $\obsp$ is no more represented by the symmetric tensor $\mathcal{R}_{\stsp}$, however, it is interesting to notice that $\mathcal{R}_{\stsp}$ introduces a measure of nonclassicality of the state in connection with observables $\mathbf{a}$ and $\mathbf{b}$.
Indeed, it measures how much the local pointwise product differs from the quantum nonlocal product when evaluated on $\rho$.
From the statistical point of view, $\mathcal{R}_{\stsp}(de_{\mathbf{a}},de_{\mathbf{a}})$ gives the covariance of the observables $\mathbf{a}$ and $\mathbf{b}$ on the state $\rho$.


The ``new'' Hamiltonian and gradient vector fields, are defined as:

\be
X_{f}:=\Lambda_{\stsp}\left(\mathrm{d}f\,,\cdot\right) \;\; \mbox{ and } \;\;\;  Y_{f}:=\frac{1}{\lambda}\mathcal{R}_{\stsp}\left(\mathrm{d}f\,,\cdot\right)\,.
\ee
Direct calculations show that:

\be
\left[Y_{e_{\mathbf{a}}}\,;Y_{e_{\mathbf{b}}}\right]=-X_{\{e_{\mathbf{a}}\,;e_{\mathbf{b}}\}} \,,
\ee
\be
\left[X_{e_{\mathbf{a}}}\,;X_{e_{\mathbf{b}}}\right]=X_{\{e_{\mathbf{a}}\,;e_{\mathbf{b}}\}} \,,
\ee
\be
\left[X_{e_{\mathbf{a}}}\,;Y_{e_{\mathbf{b}}}\right]=Y_{\{e_{\mathbf{a}}\,;e_{\mathbf{b}}\}}\,.
\ee
and thus we find that, for a quantum system with $N$ levels, the Hamiltonian and gradient vector fields defined above provide an action of the complexification $\mathfrak{sl}(N,\mathbb{C})$ of the Lie algebra $\mathfrak{su}(N,\mathbb{C})$.

With a little patience, it is possible compute the commutators of the ``new'' Hamiltonian, or gradient, vector fields with the ``old'' Hamiltonian and gradient vector fields, and see that the ``new'' Hamiltonian and gradient vector fields are tangent to the integral submanifolds of the distribution generated by the ``old'' Hamiltonian and gradient vector fields.
This means that the flow of the ``new'' Hamiltonian and gradient vector fields maps positive elements of rank $k$ in $\mathcal{O}$ into positive elements of rank $k$ in $\mathcal{O}$, that is, ``new'' Hamiltonian and gradient vector fields are tangent to $\stsp^{k}$ for all $k$.

\begin{remark}
We can represent the Lie-Jordan algebra $\obsp$ using brackets on expectation value functions $e_{\mathbf{a}}$ on $\mathcal{O}$.
Specifically, the Lie product is represented by means of the bracket associated to $\Lambda_{\stsp}$:

\be
\left\{e_{\mathbf{a}}\,,e_{\mathbf{b}}\right\}_{\stsp}:=\Lambda_{\stsp}(\mathrm{d}e_{\mathbf{a}},\mathrm{d}e_{\mathbf{b}})=e_{[[\mathbf{a},\mathbf{b}]]}\,,
\ee
while the Jordan product is represented by means of the following bracket:

\be
\left(e_{\mathbf{a}}\,,e_{\mathbf{b}}\right)_{\stsp}:=\mathcal{R}_{\stsp}(\mathrm{d}e_{\mathbf{a}},\mathrm{d}e_{\mathbf{b}}) + e_{\mathbf{a}}\,e_{\mathbf{b}}\,.
\ee
The Jordan bracket $\left(\,,\right)_{\stsp}$ is not associated to a bivector on $\mathcal{O}$, indeed, its action on the constant function $e_{\mathbb{I}}$ is not zero.
However, we can think of it as defining a bidifferential operator on $\mathcal{O}$, i.e., a bivector plus a homogeneous part, so that we obtain the first order differential operator:

\be
D_{\mathbf{a}}(\cdot)=\left(e_{\mathbf{a}}\,,\cdot\right)_{\stsp}=\mathcal{R}_{\stsp}(e_{\mathbf{a}}\,,\cdot) + e_{\mathbf{a}}\cdot\,.
\ee
\end{remark}

\begin{example}[Two-level quantum system]\label{example: two level quantum system geometry}

To illustrate our general arguments we consider the example of the two-level quantum system.
The $C^{*}$-algebra $\calg$ is isomorphic to $\mathcal{B}(\mathcal{H})$ for $\mathcal{H}\equiv\mathbb{C}^{2}$.
Thus the space $\obsp$ of the real elements is generated by the Hermitean $(2\times 2)$ Pauli matrices:

\be
\sigma_{0}=\left(\begin{array}{cc} 1 & 0 \\ 0 & 1 \end{array}\right) \;\;\;\;\;\;\;\;\;\; \sigma_{1}=\left(\begin{array}{cc} 0 & 1 \\ 1 & 0 \end{array}\right) \;\;\;\;\;\;\;\;\;\; \sigma_{2}=\left(\begin{array}{cc} 0 & -\imath \\ \imath & 0 \end{array}\right) \;\;\;\;\;\;\;\;\;\; \sigma_{3}=\left(\begin{array}{cc} 1 & 0 \\ 0 & -1 \end{array}\right)\,.
\ee
Associated to the basis $\{\sigma_{0}\,;\sigma_{1}\,;\sigma_{2}\,;\sigma_{3}\}$ there is an isomorphism $\obsp^{*}\cong\obsp$, and a Cartesian coordinate system $\{x_{0}\,;x_{1}\,;x_{2}\,;x_{3}\}$ on $\obsp^{*}$ by means of which the representative  $\rho$ of  a state is written as:

\be
\rho=\frac{1}{2}\left(\sigma_{0} + \vec{x}\cdot\vec{\sigma}\right)
\ee
with $x_{1}^{2} + x_{2}^{2} + x_{3}^{2}=1$.
In this case, $\stsp$ has only two strata, namely, $\stsp^{1}$ and $\stsp^{2}$, and it is a proper manifold with boundary\footnote{As shown in \cite{grabowski_kus_marmo-geometry_of_quantum_systems_density_states_and_entanglement} this is the only case in which $\stsp$ is a differential manifold with a smooth boundary.}.
Specifically, $\stsp$ is the $3$-dimensional solid ball and the two strata are:

\begin{itemize}
\item the pure states $x_{1}^{2} + x_{2}^{2} + x_{3}^{2}=1$;
\item the mixed states $x_{1}^{2} + x_{2}^{2} + x_{3}^{2}<1$.
\end{itemize}

It should be noticed that while pure states are represented by a compact manifold without boundary, the stratum of mixed states is bounded but not compact, and its closure is the whole of the ball.

On the affine subspace $x_{0}=1$, the expressions for $\mathcal{R}_{\stsp}$ and $\Lambda_{\stsp}$ are:

\be
\mathcal{R}_{\stsp}=\sum_{j=1}^{3}\,\frac{\partial}{\partial x_{j}}\otimes\frac{\partial}{\partial x_{j}} - \Delta_{\stsp}\otimes\Delta_{\stsp}\;\;\;\;\;\mbox{ with }\; \Delta_{\stsp}=\sum_{j=1}^{3}x_{j}\frac{\partial}{\partial x_{j}}\,,
\ee
\be
\Lambda_{\stsp}=\sum_{j,k,l=1}^{3}\epsilon_{jkl}x_{l}\frac{\partial}{\partial x_{j}}\wedge\frac{\partial}{\partial x_{k}}\,.
\ee
Gradient vector fields associated with linear functions are:

\be
Y_{j}=\frac{\partial}{\partial x_{j}} - x_{j}\,\Delta_{\stsp}\,,
\ee
while Hamiltonian ones read:

\be
X_{j}=\epsilon_{jkl}x_{l}\,\frac{\partial}{\partial x_{k}}\,.
\ee
They are complete vector fields, and together they close the Lie algebra of $SL(2,\mathbb{C})$.
Furthermore, the Hamiltonian ones are tangent to the sphere of radius $r$ for all $r>0$:

\be
\mathcal{L}_{X_{j}}r^{2}=\epsilon_{jkl}x_{l}x_{k}=0\,,
\ee
while the gradient ones are tangent only to the sphere of radius $r=1$ (the pure states):

\be
\mathcal{L}_{Y_{j}}r^{2}=2x_{j} - x_{j} (2r^{2})=2(1-r^{2})x_{j} \;\;\Longrightarrow \; \left(\mathcal{L}_{Y_{j}}r^{2}\right)\left|_{r^{2}=1}\right.=0\,.
\ee

\end{example}

\subsection{Kossakowski-Lindblad vector fields}

The algebraic description of quantum systems associated to the $C^{*}$-algebraic formulation has led to an interesting formalization of the dynamics of open quantum systems in terms of the so-called Kossakowski-Lindblad generator $\mathbf{L}$  (\cite{gorini_kossakowski_sudarshan-completely_positive_dynamical_semigroups_of_N-level_systems, lindblad-on_the_generators_of_quantum_dynamical_semigroups}, and \cite{hille_phillips-functional_analysis_and_semigroups} chapter XXV).
Specifically, a Markovian dynamics is described by the equation of motion: 

\be
\frac{d}{dt}\rho=L(\rho) \;\;\;\;\;\; \rho(t=0)=\rho_{0}\,.
\ee
According to \cite{gorini_kossakowski_sudarshan-completely_positive_dynamical_semigroups_of_N-level_systems} and \cite{lindblad-on_the_generators_of_quantum_dynamical_semigroups}, the  expression for the linear generator $\mathbf{L}$ reads:

\be\label{eqn: K-L equation}
\mathbf{L}(\rho)=-\imath\left[\mathbf{H},\rho\right] - \frac{1}{2}\sum_{j=1}^{r\leq n^{2}-1}\left\{\mathbf{V}_{j}^{\dagger}\mathbf{V}_{j},\rho\right\} + \sum_{j=1}^{r\leq n^{2}-1}\mathbf{V}_{j}\,\rho\, \mathbf{V}_{j}^{\dagger}\,,
\ee
where $\rho\in\stsp$, $\mathbf{H}^{\dagger}=\mathbf{H}$, $\mathbf{V}_{j}$ are arbitrary elements of $\mathcal{B}(\mathcal{H})$, and $\{\,,\}$ denotes the anti-commutator.
The integration of the equation of motion gives a one-parameter semigroup $\{\Phi_{\tau}\}$ of completely-positive maps $\Phi_{\tau}:\stsp\rightarrow\stsp$ for $\tau\geq0$, such that $\Phi_{0}$ is the identity transformation.
Actually, $\{\Phi_{\tau}\}$ is well-defined and differentiable for all $\tau\in\mathbb{R}$ on the whole $\obsp^{*}$, but, for $\tau<0$ it fails to preserve positivity, hence, it maps states out of $\stsp$.

Recall that a differentiable curve $\gamma:\mathbb{R}\rightarrow\obsp^{*}$ lying entirely in the full space of quantum states $\stsp$ must remain in the stratum $\stsp^{k}$ it actually belongs to, hence, the curve $\Phi_{\tau}$ lies in some $\stsp^{k}$ for all $\tau>0$.
However, when $\{\Phi_{\tau}\}$ is a semigroup, it is possible that $rank(\Phi_{0}(\rho))\leq rank(\Phi_{\tau}(\rho))$ when $\tau>0$, that is, $\{\Phi_{\tau}\}$ can ``move'' states across different strata.
This particular feature makes $\{\Phi_{\tau}\}$ a good candidate for the description of physical phenomena such a dechoerence, where the rank of the state is not preserved.

We would like to describe such a dynamics in geometrical terms, that is, using vector fields.
Since $L$ is a linear map, there is a linear vector field $\Gamma_{L}$ naturally associated with it.
Furthermore, we can write:

$$
\Gamma_{L}=\widetilde{X}_{H} - \widetilde{Y}_{V} + \widetilde{Z}_{K}\,,
$$
where $\widetilde{X}_{H}$ is the Hamiltonian vector field associated with $\mathbf{H}$ by means of $\Lambda$, $\widetilde{Y}_{V}$ is the gradient vector field  associated with $\mathbf{V}:=\frac{1}{2}\sum_{j=1}^{r\leq n^{2}-1}\,\mathbf{V}_{j}^{\dagger}\mathbf{V}_{j}$  by means of $\mathcal{R}$, and $\widetilde{Z}_{K}$ is the linear vector field associated with the linear map $\sum_{j=1}^{r\leq n^{2}-1}\mathbf{V}_{j}\,\rho\, \mathbf{V}_{j}^{\dagger}$.

Although very simple, this decomposition of the vector field $\Gamma_{L}$ is not adapted to the geometry of the space of quantum states.
Indeed, both $\widetilde{Y}_{V}$ and $\widetilde{Z}_{K}$ do not separately preserve the affine subspace $f_{\mathbb{I}}(\xi)=1$ of which the space of quantum states $\stsp$ is a subset.
Accordigly, we would like to use the Hamiltonian and gradient vector fields associated with $\Lambda_{\stsp}$ and $\mathcal{R}_{\stsp}$ in order to describe the Kossakowski-Lindblad generator $\mathbf{L}$.
However, we can not use only Hamiltonian or gradient vector fields tangent to $\stsp^{k}$ to accomplish this task.
Indeed, gradient vector fields are, in general, non-linear, while the generator $\mathbf{L}$ is linear.
Moreover, both Hamiltonian and gradient vector fields, being tangent to $\stsp^{k}$, are not able to move a state across different strata.



It turns out that is possible to define a vector field $Z_{K}$ implementing the motion across strata.
This vector field $Z_{K}$, called Kraus vector field, is defined requiring that $Z_{K}(f_{\mathbb{I}})=0$, and requiring its action on expectation value functions to be:

\be
\left(Z_{K}(e_{a})\right)(\rho)=e_{K^{\sharp}(a)}(\rho) - e_{V}(\rho)e_{a}(\rho)\,,
\ee
where $K^{\sharp}(\mathbf{a})$ is the action on $\obsp$ of the adjoint map of $K(\xi)=\sum_{j}\mathbf{V}\xi\mathbf{V}^{\dagger}$, and $\mathbf{V}=\sum_{j}\mathbf{V}_{j}^{\dagger}\mathbf{V}_{j}$.

Now, the Kossakowski-Lindblad dynamics can be described using a fine-tuned combination 

\be
\Gamma=X_{H} + Y_{G} + Z_{K}
\ee
of a Hamiltonian vector field $X_{H}$ with $\mathbf{H}\in\obsp$, a gradient vector field $Y_{G}$ with $\mathbf{G}\in\obsp$, and a Kraus vector field $Z_{K}$.
The action of a generic $\Gamma$ on the expectation value functions reads:

\be
\left(\Gamma(e_{\mathbf{a}})\right)(\rho)=e_{[[\mathbf{H}\,;\mathbf{a}]]}(\rho) + e_{\mathbf{G}\odot\mathbf{a}}(\rho) + e_{K^{\sharp}(\mathbf{a})}(\rho) -e_{\mathbf{G}}(\rho)\,e_{\mathbf{a}}(\rho) - e_{V}(\rho)\,e_{\mathbf{a}}(\rho)\,.
\ee
As it stands, $\Gamma$ does not describe the Kossakowski-Lindblad dynamics since it is not linear.
The non-linearity of the vector field $\Gamma$ is due to the last two terms, hence, choosing $\mathbf{G}=-\mathbf{V}$ the resulting vector field is linear.
In conclusion, the the Kossakowski-Lindblad dynamics of a quantum system is described by the Kossakowski-Lindblad vector field:

\be
\Gamma_{L}=X_{H} + Y_{G} + Z_{K}\,,
\ee
with $\mathbf{G}=-\mathbf{V}$.
Note that, once the decomposition of the Kraus map $K$ is given, the vector field $Y_{G}$ is automatically defined by the condition $\mathbf{G}=-\mathbf{V}$, furthermore, $Z_{K}$ is the vector field generating the transversal motion which  to the strata of $\stsp$ characterizing dissipative phenomena.
From the point of view of $\stsp$, the Kossakowski-Lindblad vector field generates a one-parameter semigroup of transformations describing dissipative quantum dynamics.

\begin{example}[Phase damping of a qubit]

Let us consider the Kossakowski-Lindblad dynamics generated by equation (\ref{eqn: K-L equation}) with $\mathbf{H}=0$, and $\mathbf{V}_{1}=\sqrt{1-\gamma}\,\mathbb{I}$, $\mathbf{V}_{2}=\sqrt{\gamma}\,\sigma_{3}$:

\be
\mathbf{L}(\rho)=-\gamma\left(\rho - \sigma_{3}\,\rho\,\sigma_{3}\right)\,.
\ee
This dynamical evolution is known as phase damping.
Indeed, with the notation of example \ref{example: two level quantum system geometry}, the explicit form of the  dynamics generated by $\mathbf{L}$ reads:

\be
\Phi_{\tau}(\rho)=\frac{1}{2}\left(\sigma_{0} + \exp(-\gamma\tau)\left(\rho^{1}\sigma_{1} + \rho^{2}\sigma_{2}\right) + \rho^{3}\sigma_{3}\right)\,,
\ee
and it is clear that this dynamics only affect the phase terms (off-diagonal terms) of $\rho$ represented by the components along $\sigma_{1}$ and $\sigma_{2}$.

The Kossakowski-Lindblad vector field on the affine subspace $f_{\mathbb{I}}=1$ can easily be calculated.
At this purpouse, note that $\mathbf{H}=0$ implies $X_{H}=0$, and, since $-\mathbf{G}=\mathbf{V}=\sum_{j}\mathbf{V}_{j}^{\dagger}\,\mathbf{V}_{j}=\mathbb{I}$, it is $Y_{G}=0$.
Consequently, the Kossakowski-Lindblad vector field $\Gamma_{L}$ reduces to the Kraus vector field $Z_{K}$, and the final expression is:

\be
\Gamma_{L}=-2\gamma\left(x_{1}\,\frac{\partial}{\partial x_{1}} + x_{2}\,\frac{\partial}{\partial x_{2}}\right)\,.
\ee
The integral curves of $\Gamma_{L}$ are radial lines lying in two-dimensional planes orthogonal to the $x_{3}$-axis.
Consequently, we can see that, if the initial state $\rho$ is a mixed state, the flow of $\Gamma_{L}$ will not take $\rho$ out of $\stsp^{2}$ for every $\tau\geq0$, however, since it is transversal to the spheres centered in $x_{1}=x_{2}=x_{3}=0$ representing isospectral states, it will change the spectrum of $\rho$ giving rise to dissipation.
If the initial state $\rho$ is a pure state, the flow of $\Gamma_{L}$ will take $\rho$ from $\stsp^{1}$ to $\stsp^{2}$ right after $\tau=0$, giving rise again to a dissipative behaviour.
Note that, for $\tau\leq0$, the flow of $\Gamma_{L}$ takes a state $\rho$ out of $\stsp$, therefore, from the point of view of the space of states $\stsp$, $\Gamma_{L}$ generates a one-parameter semigroup of transformations.
\end{example}

\section{Conclusions}

In many physical situations, the carrier spaces we are interested in do not possess a ``canonical'' differential structure, hence, there is the need to develop methods to handle such special situations.
Unfortunately, the mathematical tools present in the literature do not always capture  relevant physical aspects of some interesting situations, as it is the case, for example, for the one-dimensional bouncing particle described  in section \ref{section: introduction}.

In this paper we reviewed some alternative methods to deal with carrier spaces presenting no global differential structure.
The main tool we adopted is the ``duality relation'' existing between a set $S$ and the algebra of functions $\mathcal{F}(S)$ on it.
Different mathematical structures defined on $S$ can be directly encoded in the choice of a suitable algebra of function $\mathcal{F}(S)$, e.g., for a smooth manifold $M$, $\mathcal{F}(M)$ is the algebra of smooth functions.
The advantage of working with the space of functions $\mathcal{F}(S)$ rather than with $S$ is that, on $\mathcal{F}(S)$, there is an algebra structure at our disposal.
Furthermore, the use of $\mathcal{F}(S)$ perfectly fits into the so-called unfolding-reduction scheme (see, for instance \cite{carinena_ibort_marmo_morandi-geometry_from_dynamics_classical_and_quantum} chap. $7$), according to which it is possible to overcome some of the technical difficulties associated to non-trivial carrier spaces by unfolding them into nice, and possibly linear, ambient spaces.
For example, if the reduction is associated with a quotienting procedure, then the algebra of functions of the non-trivial carrier spaces becomes a subalgebra of the algebra of functions of the ambient space.

Particular attention has been devoted to the case of manifolds with boundary.
Specifically, we focused on the concrete example of a finite-level quantum system associated with a finite-dimensional $C^{*}$-algebra $\mathcal{A}$.
In this case, the carrier space is the space $\stsp$ of quantum states which is a stratified manifold with boundary.
A suitable ambient space for $\stsp$ is the dual space $\obsp^{*}$ of the space $\obsp$ of observables of the theory, which is a linear space.
According with the philosophy exposed in the paper, we represent the algebra $\mathcal{F}(\stsp)$ of functions on the space of quantum states, in the algebra $\mathcal{F}(\obsp^{*})$ of smooth functions on $\obsp^{*}$.
This procedure enables us to define a vector field $\Gamma_{L}$ on $\obsp^{*}$ (a derivation of $\mathcal{F}(\obsp^{*})$ in the algebraic language) which is transversal to the strata of $\stsp$ and is the generator of a semigroup (possibly a local one).
By means of the vector field $\Gamma_{L}$ it is possible to describe physical processes, such as decoherence and damping, in which the rank of quantum states is not preserved.
Furthermore, $\Gamma_{L}$ leads to a semigroup of transformations on $\stsp$, and thus, this physical example suggests that derivations generating groups of transformations are not enough to fully describe physically interesting situations, and we need to consider generators of semigroups of transformations.

\addcontentsline{toc}{section}{References}
\bibliographystyle{unsrt}
\bibliography{scientific_bibliography}

\end{document}